\documentclass[epj]{svjour}
\usepackage{graphicx}
\setkeys{Gin}{width=0.4\textwidth}

\begin{document}
\title{Small Bipolarons in the 2-dimensional Holstein-Hubbard Model}
\subtitle{II  Quantum Bipolarons}
\author{L. Proville  \and S. Aubry}
\institute{DAMTP Cambridge
University, Cambridge, CB3 9EW, UK\\
Laboratoire L\'eon Brillouin (CEA-CNRS), CEA Saclay
	91191-Gif-sur-Yvette Cedex, France}

\date{Accepted  Ref. B9602 }

\abstract{We study the effective mass of the bipolarons and
essentially  the possibility to get both light and strongly bound
 bipolarons in the Holstein-Hubbard model and some variations
in the vicinity of the adiabatic limit.
Several approaches to investigate the quantum mobility of
polarons and bipolarons are proposed for this model.
First, the quantum fluctuations
are treated as perturbations of the mean-field (or
adiabatic) approximation
of the electron-phonon coupling in order to calculate the bipolaron bands.
It is found that the bipolaron mass generally remains very large
except in the vicinity of the triple point of the phase diagram (see
\cite{PA99}),
where the bipolarons have several degenerate configurations at the
adiabatic limit
(single site (S0), two sites (S1) and quadrisinglet (QS)),  while the
polarons are much lighter.
This degeneracy reduces  the bipolaron mass significantly. Next we
improve this result
by variational methods (modified Toyozawa Exponential  Ansatz or TEA)
valid for larger quantum perturbations away from the adiabatic limit.
We first test this new method for the single polaron.
We find that the triple point of the phase diagram is washed out by the
lattice quantum fluctuations which thus suppress the light bipolarons.
Further improvements of the method by hybridization of several TEA states
do not change this conclusion.
Next we show that some  model variations , for example
a phonon dispersion
may increase the stability of the (QS)
bipolaron against the quantum lattice fluctuations.
We show that the triple point of the phase
diagram may be stable
to quantum lattice fluctuations and a very sharp mass reduction may
occur, leading to bipolaron masses of the
order of 100 bare electronic mass for realistic parameters.
Thus we argue that such very light bipolarons could condense as a
superconducting
state at relatively high temperature when
their interactions are not too large,
that is, their density is small enough. This effect might be relevant for
understanding
the origin of the high Tc superconductivity of doped cuprates far enough
from half filling.
}
\maketitle

\section{Introduction}

\subsection{Specific Problem for high TC Superconductivity}

Superconducting materials at temperatures significantly higher than the
maximum Tc predicted by MacMillan \cite{McM68} for the standard BCS
superconductivity \cite{BCS57} are exceptional \cite{BM86}. Up to now,
there is a wide variety of
such materials which are all cuprates built with  CuO$_{2}$ planes
and with many kinds of interlayer dopants \cite{Wal96}.

When  the electron-phonon coupling increases too much, it is known for
several decades (Migdal \cite{Migd58}) that the BCS theory should break
down because of lattice instabilities. These instabilities are associated
with the formation of polarons and bipolarons. Alexandrov et al.
\cite{ARR86} developed later a theory of bipolaronic
superconductivity where the electrons form strongly bound on-site bipolarons.
They are described by a hard core boson model which could become superfluid.
Unfortunately, their calculations also show that when the electron phonon
coupling increases beyond the Migdal instability, the effective mass of
these bosons grows exponentially fast and becomes so huge that it seems
hopeless to get superconductivity in this model, at least at non-negligible
temperatures.

One of us(SA) conjectured in \cite{Aub93,Aub95b} that the
interplay of
the electron-phonon coupling with a direct electron-electron repulsion
 could reduce significantly the bipolaron effective mass and thus
favor high Tc superconductivity. In the present paper we support this
 conjecture by calculating the effective mass of a single bipolaron
in a 2d model which involves both an
electron-phonon coupling and a direct electron-electron repulsion.
For this  purpose  we choose to investigate first this effect in
the Holstein-Hubbard model because of its simplicity.

In the absence of Hubbard repulsion, we confirm that the
effective mass of the bipolaron is indeed very large \cite{CRF98}, which is
incompatible with a high Tc superconducting phase.
When the Hubbard term is increased new bipolaronic states
become stable. They are 2-site bipolarons which consist of
two neighboring polarons bound by their magnetic interaction in a
singlet state and also a bipolaron
called a ``quadrisinglet'' (QS) which consists of the combination of four
singlets sharing one common site. In the parameter region where
these bipolaronic states have nearly degenerate energies,
the effective
bipolaron mass is sharply reduced.
The drastic mass reduction is due to resonance between
the different bipolarons. Certain variations of the model, such as a
phonon dispersion, might even increase the binding energy
of the bipolaron while allowing it to keep
a very light  effective mass for realistic parameters.

We first discuss some early known results about the effective
masses of polarons and bipolarons.
Adiabatic results (section \ref{sect2}) described in \cite{PA99}
are briefly recalled. Next
we treat the quantum lattice fluctuations as a perturbation of the
adiabatic limit. The main effect is to lift the
bipolaron degeneracy both due to translational invariance  and to the
possible existence of several kinds of adiabatic bipolarons with almost the
same energy (section \ref{BBW}). This correction is only valid for
a very small quantum lattice parameter.

To extend the field of validity of our calculations we next propose to
use the Toyozawa variational form. Quantum
polarons and bipolarons are approximated by
a self-consistent Bloch wave
that is exact at the adiabatic limit.
At this limit it is demonstrated in \cite{PA99}
that there is not a great loss of accuracy if
the shapes of polaron and bipolaron are exponentials.
Thanks to this approximation we gain much
simplicity for the variational form.
We first apply this method to polarons (section \ref{sect4}) and next to
bipolarons (section \ref{sect5}). For very small quantum lattice
fluctuations the results obtained by
perturbation of the adiabatic limit
 are practically recorvered but there are significant deviations when the
fluctuations become larger. First order transitions that cannot exist
physically are washed out by hybridizing several Toyozawa
variational forms. Actually, the regime where the bipolaron mass is sharply
reduced is swallowed up in the domain where the ground-state is
unbound polarons.
We demonstrate in section (\ref{sect6}) that this undesirable phenomena
can be avoided
by variations ofthe  model which  increase the stability of the (QS)
bipolaron (e.g. a phonon dispersion with the
appropriate sign).

\subsection{The Holstein-Hubbard Model}

Let us first recall our notations for the model we study here. Its
Hamiltonian is
\begin{eqnarray}
\mathcal{H}&=& -T \sum_{<j,k>,\sigma} C_{j,\sigma}^{+}C_{k,\sigma}
+\sum_{j} \hbar \omega_0(a_{j}^{+}a_{j}+\frac{1}{2})\nonumber\\
&& \qquad +g n_{j}(a^{+}_{j}+a_j)
+{\upsilon}n_{j,\uparrow}n_{j,\downarrow} \label{hamiltonian}
\end{eqnarray}

where $j$ and $k$ represent lattice sites, $T$ is the transfer integral
between nearest neighbor sites $<j,k>$.  The electrons are Fermions represented
by the standard anti-commuting operators $C_{j,\sigma}^{+}$  and $C_{j,\sigma}$
at site $j$ with spin $\sigma = \uparrow$ or $\downarrow$. $n_{i,\sigma}
=C_{j,\sigma}^{+} C_{j,\sigma}$ and  $n_{i}=n_{i\uparrow}+n_{i\downarrow}$.
$a^{+}_j$ and $a_j$ are standard creation and annihilation boson
 operators of phonons and $\hbar \omega_0$ is the phonon energy of
 a dispersion-less optical phonon branch. $g$ is the on-site electron
 phonon coupling and $\upsilon$ the on-site electron-electron
 repulsion (Hubbard interaction).

 We choose $E_0=8 g^2/\hbar\omega_0$ as energy unit as in \cite{PA99}.
 Defining the position and momentum operators as
 \begin{eqnarray}
	 u_j &=& \frac{\hbar\omega_0}{4g}(a_{j}^{+}+a_{j})
	\label{position}  \\
	p_{j} &=& i\frac{2g}{\hbar\omega_0}(a_{j}^{+}-a_{j})
	\label{momenta}\\
	 &&
	\mbox{with the commutation relation} \ [u_{j},p_{j}] = i
 \end{eqnarray}

we obtain the dimensionless Hamiltonian
\begin{eqnarray}
H &=& \sum_{j} {\frac{1}{2} (u_j^2 + u_j n_j)
 + U n_{j\uparrow}n_{j\downarrow}}\nonumber\\
 && \qquad -\frac{t}{2}\sum_{<j,k>,\sigma} C_{j,\sigma}^{+}C_{k,\sigma}
 + \sum_{j} \frac{\gamma}{2} p_j^2 \label{hamduc}
\end{eqnarray}

Our reduced dimensionless parameters are
\begin{eqnarray}
 &&  E_0=8 g^2/\hbar\omega_0 \qquad   U = \frac{\upsilon}{E_{0}} \nonumber\\
 &&  t =  \frac{T}{E_{0}} \qquad
	 \gamma = \alpha^2 = \frac{1}{4} (\frac{\hbar\omega_0}{2g})^4
\label{param}
\end{eqnarray}

Despite the primitive nature of our model, it may catch important aspects of
reality.
However, we shall also demonstrate at the end of this paper that
certain model variations could be favorable for bipolaron mass reduction.
For example, we may introduce a coupling between nearest
neighbor atoms so that the phonon branch is no longer
dispersionless.
Then the new Hamiltonian is  the sum of (\ref{hamduc}) and the extra energy
term
$-C \sum_{<i,j>} u_{i}u_{j}$ where $<i,j>$ represents all the pairs of nearest
neighbor sites $i$ and $j$. When $C>0$, the bipolaron
mass reduction
is enhanced while it remains strongly bound.
Thus we demonstrate that relatively minor changes in the model may favor
(or disfavor)  superconductivity at relatively high temperature.

\subsection{Polaron and Bipolaron Effective Mass}

Let us first briefly recall some standard results about the
effective masses of polarons and bipolarons. The Lang-Firsov unitary
transformation
\cite{Lang_Firsov} yields  a new Hamiltonian
\begin{eqnarray}
	H_{LF}&=&\sum_{j} {\frac{1}{2} u_j^2 + U n_{j\uparrow}n_{j\downarrow}}
	- \frac{1}{8}(n_{j\uparrow}+n_{j\downarrow})^{2}  \nonumber \\
 &-& \frac{t}{2}\sum_{<j,k>,\sigma} e^{-i(p_{j}-p_{k})/2}
 \tilde{C}_{j,\sigma}^{+}\tilde{C}_{k,\sigma}
 +\frac{\gamma}{2} \sum_{j} p_j^2
	\label{hamiltLF}
\end{eqnarray}
where  $H_{LF}=e^{-iS_{LF}} H e^{iS_{LF}}$ with $S_{LF}=\frac{1}{2}\sum_{j}
{ p_{j} n_{j}}$

After this transformation, the creation operator $\tilde{C}_{j,\sigma}^{+}$
at site $j$ acts on the vacuum by creating both
an electron and  a lattice distortion
\begin{equation}
	\tilde{C}_{j,\sigma}^{+}|\emptyset> = C_{j,\sigma}^{+} e^{-i
p_{j}/2} |\emptyset>
	\label{polcreat}
\end{equation}
that is, a polaron. A standard but rough mean-field  approximation consists in
taking the average of the transfer integral for  unperturbed phonons. We
obtain an
approximate formula for the transfer integral of this polaron
 \begin{equation}
 T_{LF} = t e^{-1/(8 \alpha)}
	\label{poltrint}
 \end{equation}

 When there is a single electron in the system the electron-electron
interaction
 does not play  any role. Then the effective mass of a single polaron is
 defined as the inverse of
 the second order derivative versus the wave-vector $q$ of
 the polaronic energy
 $E(q)$, that is, $T_{LF}$.
 The effective mass of the polaron is that of the bare
 electron multiplied by $\exp{[1/(8 \alpha)]}$. It
  becomes exponentially large when $\alpha$ is small.

 This approximation tends to become right only when the operator corresponding
to the transfer
 integral in eq.\ref{hamiltLF} has small fluctuations. This condition
 is fulfilled when the pre-factor $t$ is small. It is also fulfilled when
 $\alpha$ is large, that is,
 for a weak electron-phonon coupling $g$ compared to the phonon energy
 $\hbar \omega_{0}$.
 In the antiadiabatic limit ($\alpha$ large), the model becomes a Hubbard model
 with an on-site effective electron-electron  interaction $\tilde{U}=(U-1/4)$
 which is attractive for $U<1/4$ and repulsive for $U>1/4$ and where
 the transfer integral has been renormalized.
 The negative $U$ model is expected to have superconducting phases
 \cite{NS85} for non-vanishing band filling. However, we treat here the
opposite case
 $\alpha$ small which is close to the adiabatic limit.

 Actually when $t$ is small, our numerical results
 agree with formula (\ref{poltrint}). For larger $t$,
 the effective mass
 of the polaron  given by (\ref{poltrint}) becomes larger  than the
 mass we compute. Note that our result should be more reliable
 because  it yields a lower variational energy for a single polaron.

 The effective mass of the bipolaron has been calculated by Alexandrov et al
 in the same limit ($t$ small) \cite{ARR86} for strongly bound bipolarons
(that is, for $U$ small)
and considering the kinetic energy term in eq.\ref{hamiltLF}
 as a perturbation. In our dimensionless units they found the transfer
integral $t_{b}$ for a bipolaron
\begin{equation}
	t_{b} = \frac{4 t^{2}}{1-4U} e^{-1/(2 \alpha)}
	\label{bipoltrint}
\end{equation}

If one extrapolates naively this formula for larger $U$, one would find that
$t_b$ becomes infinite. Of course,
the associated effective mass of the
bipolaron cannot
vanish, but  our results nevertheless demonstrate that it is sharply depressed
not far from
the region $U \approx 0.25$.
Comparison of formula (\ref{poltrint}) and \ref{bipoltrint}
shows that in
the region where both $\alpha$ and $t$ are small
the effective mass, of the bipolaron
is much larger by many
order of magnitude  than the polaron mass which  is itself much larger than
that of
the bare electron. In most physically realistic situations, the bipolaron
masses
are so huge that it is unreasonable to consider that they could
exhibit a Bose condensation  \cite{CRF98}.

We perform here a numerical calculation of the effective mass of the
bipolaron (and also the polaron) in order to show that in some specific
regions of the parameter space, when the Hubbard term
becomes comparable with the electron-phonon binding energy,
 these
effective masses can be drastically reduced so that Bose
condensations of bipolarons become plausible.

\section{The Mean-field Holstein-Hubbard Model}
\label{sect2}

 We calculate first the adiabatic bipolarons
 which are ground-state of a mean-field Hamiltonian. They are the exact
 solutions in the adiabatic limit when $\gamma=\alpha^{2}$ is zero
 (that is, when the atomic kinetic energy is negligible).
 These spatially localized solutions are degenerate under lattice
translation. This
 degeneracy is lifted when the atomic kinetic energy is taken
 into account. Within a perturbative treatment,
 this explicitly gives  bands of
 extended bipolarons characterized by a wave-vector. The inverse curvature
of the lowest band at zero wave-vector yields the bipolaron
effective mass. This calculation become exact in principle in the limit of
small $\gamma$.
 Note that similar methods were already developed in \cite{QCA92} to
calculate the effective masses
 of discommensurations in Charge Density Waves.

\subsection{The adiabatic regime}
The Hamiltonian eq.\ref{hamduc} can be written as the sum of three terms
\begin{equation}
 H = H_{el}+H_{ph}+H_{f}
	\label{hammf}
\end{equation}
where $H_{el}$ and $H_{ph}$ are decoupled electron and phonon Hamiltonians
respectively and $H_{f}$ is a fluctuation term.
\begin{eqnarray}
H_{el} &=& \sum_{i}{\left( \frac{1}{2} \bar{u}_i n_i+
U n_{i\uparrow}n_{i\downarrow}\right)}
 -\frac{t}{2}\sum_{<i,j>,\sigma}C_{i,\sigma}^{+}C_{j,\sigma}
 \nonumber \\  \label{hamiltel} \\
H_{ph} &=& \frac{1}{2} \sum_{i}{\left(u_i^2 + u_i \bar{n}_i
- \bar{u}_i \bar{n}_i +\gamma p_i^2 \right)}
\label{hamiltph}\\
H_{f} &=& \frac{1}{2} \sum_{i} (u_{i}-\bar{u}_{i})(n_{i}-\bar{n}_{i})
\label{fluctham}
\end{eqnarray}

$\bar{n}_{i}$ and $\bar{u}_{i}$ are variational parameters which are
determined by minimizing the ground-state energy of the effective Hamiltonian
$H_{ad}=H_{el}+H_{ph}$. It comes out that $\bar{u}_{i}=\langle
u_{i}\rangle$ and
$\bar{n}_{i}=\langle n_{i}\rangle$ are the average of the corresponding
operators.
The standard mean-field approximation for the electron phonon coupling
consists in neglecting the fluctuation energy $H_{f}$.

Minimizing the ground-state energy of Hamiltonian eq.\ref{hamiltph} also yields
\begin{equation}
\langle u_{i}\rangle =-\langle n_{i}\rangle /2
	\label{densMF}
\end{equation}

Then, the ground-state  of the mean field
Hamiltonian $H_{ad}$ has the form
\begin{equation}
	|\Psi> = \left(\sum_{i,j} \psi_{i,j} C_{i,\uparrow}^{+}
	C_{j,\downarrow}^{+}\right).  \exp \left(i \sum_{n}\bar{u}_{n}
	p_{n}\right) |\emptyset>
	\label{elecgs}
\end{equation}

A pair of electrons with the electronic wave function $\psi_{i,j}$ is
created as well as the corresponding lattice distortion
$\bar{u}_{i}$. The electronic wave function is a
singlet state, that is, a symmetric
function of $(i,j)$: $\psi_{i,j}=\psi_{j,i}$ .
It fulfills an extended nonlinear Schroe\-dinger equation
which is exactly the same as in the adiabatic case at $\alpha=0$
\cite{PA99}.
\begin{equation}
-\frac{t}{2}\Delta \psi_{i,j}
+\left(-\frac{\bar{n}_i+\bar{n}_j}{4}+U\delta_{i,j}
\right) \psi_{i,j}=E_{el} \psi_{i,j} \label{NLSE}
\end{equation}

$\Delta$ is the four-dimensional discrete Laplacian and $\bar{n}_{i}=
	\bar{n}_{i,\uparrow}+   \bar{n}_{i,\downarrow}$ with
\begin{equation}
	\bar{n}_{i,\uparrow} = \sum_{j} |\psi_{i,j}|^{2}
	\label{updensity}
\end{equation}

The square root of the mean square lattice fluctuation $\langle
(u_{i}-\bar{u}_{i})^{2}\rangle^{1/2}=\alpha={\gamma}^{1/2}$ is small of
order $\alpha$.
Thus, the mean-field approximation obviously becomes exact in the
adiabatic limit
$\alpha \rightarrow 0$ when there are no lattice fluctuations.

\subsection{Bipolarons from Anti-integrable limit and variational
approximations}
For an easy understanding the reader should refer to our
early paper \cite{PA99} were the adiabatic (or mean
field) bipolarons were investigated in detail in the
two-dimensional model by continuation from the
anti-integrable limit $(t=0)$.

The main result of \cite{PA99}
is that we found a quite rich phase
diagram with first order
transition lines in the parameter space $(U,t)$. For large $t$ the
electrons remain extended and  do not self localize as bipolarons.
For small $t$ there are several kinds of structures that compete
to be the bipolaron ground-state. These bipolarons were denoted
(S0), (S1) and (QS).  Bipolaron (S0) is mostly  localized at a single site
and has square symmetry. Bipolaron (S1) consists into a bound pair of
polarons
in a magnetic singlet state localized on two neighboring sites.
It breaks the square symmetry and is oriented either in the $x$ direction
(S1)$_{x}$ or  the $y$ direction (S1)$_{y}$. The quadrisinglet
bipolaron (QS) is a combination of four singlet states  with a common
central site and has square symmetry.

Interesting properties are obtained at a triple point
corresponding to
the intersection of three first-order transition lines. At that point,
and apart from the translational degeneracy,
there are four different degenerate bipolarons (S0), (QS),
(S1)$_{x}$
and  (S1)$_{y}$. We shall see that the
quantum lattice perturbations hybridize these
degenerate states and hence
drastically enhance the bandwidth of the
bipolaron or, equivalently, reduce its effective mass.
Within a classical picture we already noticed that the
energy barrier ( Peierls-Nabarro barrier) which has to be overcome to
move the bipolaron through the lattice was drastically reduced.

We also investigated in \cite{PA99} some approximations
with exponential variational forms for the bipolarons that allow
analytical calculations. The exact phase diagram calculated numerically
was reproduced with a quite good accuracy with the following forms
\begin{eqnarray}
\psi_{i,j}^{S0}&=&A \lambda^{(|i_x|+|i_y|+|j_x|+|j_y|)} \label{AS0}\\
\psi_{i,j}^{S1}&=&\frac{B}{\sqrt{2}}
(\lambda^{(|i_x-1|+|i_y|+|j_x|+|j_y|)}\nonumber\\
&& \qquad +\lambda^{(|i_x|+|i_y|+|j_x-1|+|j_y|)})\label{AS1}\\
\psi_{i,j}^{QS}&=&\frac{C}{\sqrt{8}} \sum_\pm
 \lambda_{2}^{(|j_x|+|j_y|)} (\lambda_{1}^{(|i_x \pm
1|+|i_y|)}+\lambda_{1}^{(|i_x|+|i_y\pm 1|)})
\nonumber\\
&& \ +\lambda_{2}^{(|i_x|+|i_y|)}
(\lambda_{1}^{(|j_x\pm 1|+|j_y|)}+\lambda_{1}^{(|j_x|+|j_y\pm 1|)} )
\label{ASQ}
\end{eqnarray}
for bipolarons (S0), (S1) and (QS) respectively. A, B, and C are
normalization factors
and the parameters $\lambda$, $\lambda_{1}$ and $\lambda_{2}$ are optimized
for energy
minimization. We shall develop here a
quantum analogous version of these approximations to improve our
methods.

% ***************************************
% ***************************************
\section{Quantum Lattice Corrections}
% ***************************************
% ***************************************

\label{BBW}
 We now treat the mean-field fluctuation
$H_{f}=1/2 \sum_{i}(u_i-\bar{u}_{i})(n_i-\bar{n}_{i})$
 as a perturbation that lifts the translational degeneracy of the
 mean field bipolarons (\ref{elecgs}), whose  wave functions
 are denoted $|\Omega^{S}(j)>$ where S represents bipolarons (S0),
 (S1)$_{x}$, (S1)$_{y}$ or (QS). The index $j$ is the site where the bipolaron
 (S) is located. For bipolarons (S1)$_{x}$ and (S1)$_{y}$
 which occupy two adjacent
 sites $(j_{x},j_{y})$ and $(j_{x}+1,j_{y})$
 or $(j_{x},j_{y})$ and $(j_{x},j_{y}+1)$, respectively,
 we choose by convention
 $j=(j_{x},j_{y})$.
 To treat the mean-field fluctuation in lowest order, the
 initial Hamiltonian (\ref{hammf}) should
 be projected and diagonalized in the subspace
generated by
 all these translated wave functions.
  We already noticed that the bipolaron energies might be
 degenerate or almost degenerate so that we should take into account
 their possible hybridization. The eigenstates should have the general form

 \begin{equation}
	|\Omega> = \sum_{S,j} a_{S,j} |\Omega^{S}(j)>
	\label{gnform}
 \end{equation}
 where $a_{S,j}$ are coefficients to be determined by extremalization
 of $<\Omega|H|\Omega>$ with the normalization constraint  $<\Omega|\Omega>=1$.
 Both $<\Omega|\Omega>$ and $<\Omega|H|\Omega>$ are quadratic
 functions of $a_{S,j}$
 \begin{eqnarray}
	<\Omega|H|\Omega> & = & \sum_{(S,i),(S^{\prime},j)} a_{S,i}^{*}
M_{(S,i),(S^{\prime},j)}
	a_{S^{\prime},j}
	\label{matH}  \\
		<\Omega|\Omega> & = & \sum_{(S,i),(S^{\prime},j)}
a_{S,i}^{*} P_{(S,i),(S^{\prime},j)}
			a_{S^{\prime},j}
	\label{matS}
 \end{eqnarray}
 where the coefficients of matrices $\textbf{P}$ and $\textbf{M}$ are
defined as
 \begin{eqnarray}
	 P_{(S,i),(S^{\prime},j)} & = &  <\Omega_{S}(i)|\Omega_{S^{\prime}}(j)>
	\label{overlap}\\
	 M_{(S,i),(S^{\prime},j)} & = &
<\Omega_{S}(i)|H|\Omega_{S^{\prime}}(j)>
	\label{Hoverlap}
 \end{eqnarray}
 It is important to take into account the fact that the eigenstates of the
 \textit{self-consistent} mean-field Hamiltonian $H_{ad}$ are not
orthogonal one with
 each other, since the matrix of scalar products is not diagonal.
 For two normalized wavefunctions $|\Omega>$ and $|\Omega^{\prime}>$ with
the form
 (\ref{elecgs}) and with  electronic wave functions $\{\psi_{i,j}\}$ and
$\{\psi_{i,j}^{\prime}\}$,
 electronic densities $\bar{n}_{i}= -2 \bar{u}_{i}$ and
 $\bar{n}_{i}^{\prime}=-2 \bar{u}_{i}^{\prime}$
 respectively the scalar products eqs.(\ref{Hoverlap}),(\ref{overlap})
 can be calculated explicitly for
 the Hamiltonian (\ref{hamduc}).
 \begin{eqnarray}
 <\Omega |\Omega^{\prime}>&=&\exp - \frac{1}{4 \alpha} \left( \sum_{i}
	(\bar{u}_{i}-\bar{u}_{i}^{\prime})^{2}\right)
 \nonumber \\  \qquad   &\times& \left(\sum_{i,j} \psi_{i,j}^{*}
\psi_{i,j}^{\prime}\right)
	\label{scpr}  \\
	<\Omega | H |\Omega^{\prime}>&=& N \frac{\alpha}{2}<\Omega
|\Omega^{\prime}> \nonumber \\
	&+& \exp - \frac{1}{4 \alpha} \left(
\sum_{i}\bar{u}_{i}-\bar{u}_{i}^{\prime})^{2}\right)
	\times \nonumber \\
	 && \qquad [ \frac{1}{2} ( \sum_{n} \bar{u}_{n}\bar{u}_{n}^{\prime})
	\times (\sum_{i,j} \psi_{i,j}^{*} \psi_{i,j}^{\prime})\nonumber \\
	&& +\frac{1}{4} \sum_{n,j} (\bar{u}_{n}+\bar{u}_{n}^{\prime})
	 \left(\psi_{n,j}^{*} \psi_{n,j}^{\prime}
	+\psi_{j,n}^{*} \psi_{j,n}^{\prime}\right)\nonumber \\
	&& + U \sum_{i} \psi_{i,i}^{*} \psi_{i,i}^{\prime}
	-\frac{t}{2} \sum_{i,j} \left(\psi_{i,j}^{*} \Delta
	\psi_{i,j}^{\prime}\right)]
	\label{hamscp}
 \end{eqnarray}
where $\Delta \psi_{i,j}= \sum_{} \psi_{k,l}$ is the discrete Laplacian on
a 4d lattice.

The extremalization equation of  $<\Omega|H|\Omega>$ with respect to
$\textbf{A}=\{a_{S,j}\}$ with the normalization condition
$\textbf{A}^{*}.\textbf{P}.\textbf{A}=1$, is $\textbf{M}.\textbf{A} = E.
\textbf{P}.\textbf{A}$
where $E$ is the Lagrange parameter, which is also the eigenenergy. It can
be written as
an eigenvalue problem for the normalized vector
$\textbf{B}=\textbf{P}^{1/2}.\textbf{A}$
\begin{equation}
	\textbf{P}^{-1/2}.\textbf{M}.\textbf{P}^{-1/2}.\textbf{B} = E.
\textbf{B}
	\label{Sch1}
\end{equation}

Note that the extensive term $N \frac{\alpha}{2}<\Omega |\Omega^{\prime}>$
in the second term
of eq.(\ref{hamscp}) does not disturb the calculations. It yields a constant
term $N \frac{\alpha}{2}$
in the effective Hamiltonian $\textbf{P}^{-1/2}.\textbf{M}.\textbf{P}^{-1/2}$
which is nothing but the
zero point phonon energy of the system with size $N$ (without electrons).

\begin{figure}
\begin{center}
\includegraphics[width=0.4\textwidth,height=0.31\textwidth]{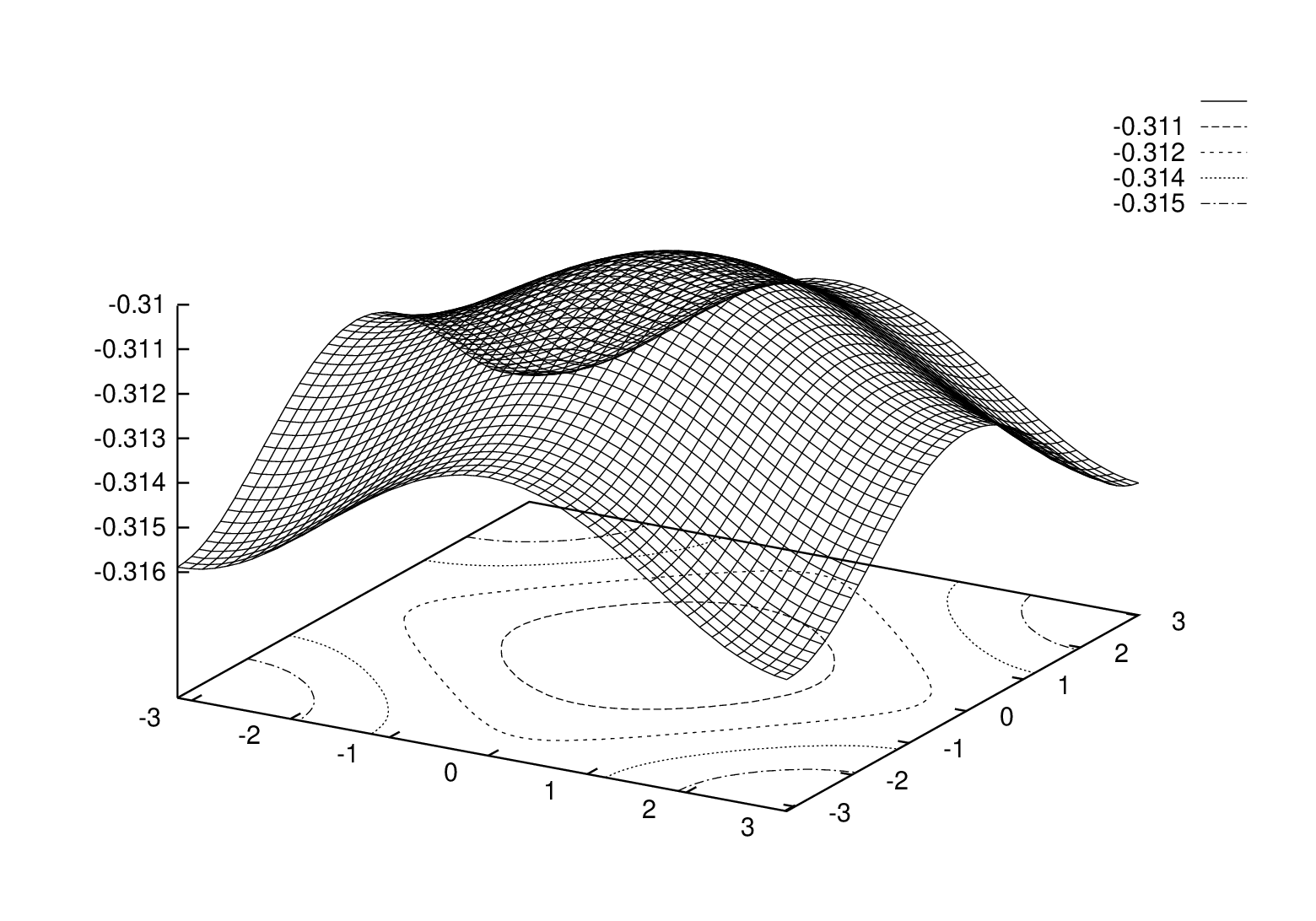}
\includegraphics[width=0.4\textwidth,height=0.31\textwidth]{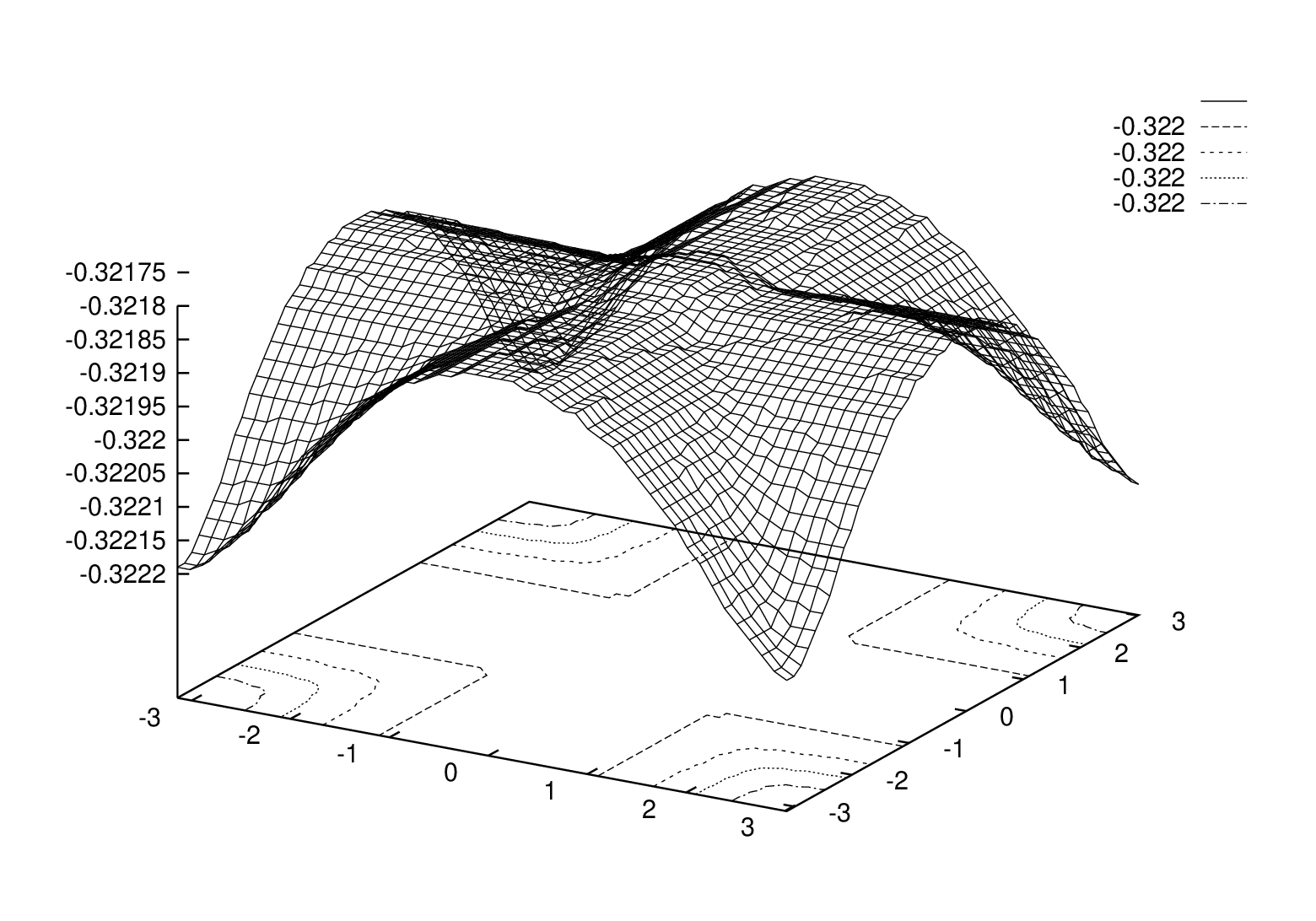}
\includegraphics[width=0.4\textwidth,height=0.31\textwidth]{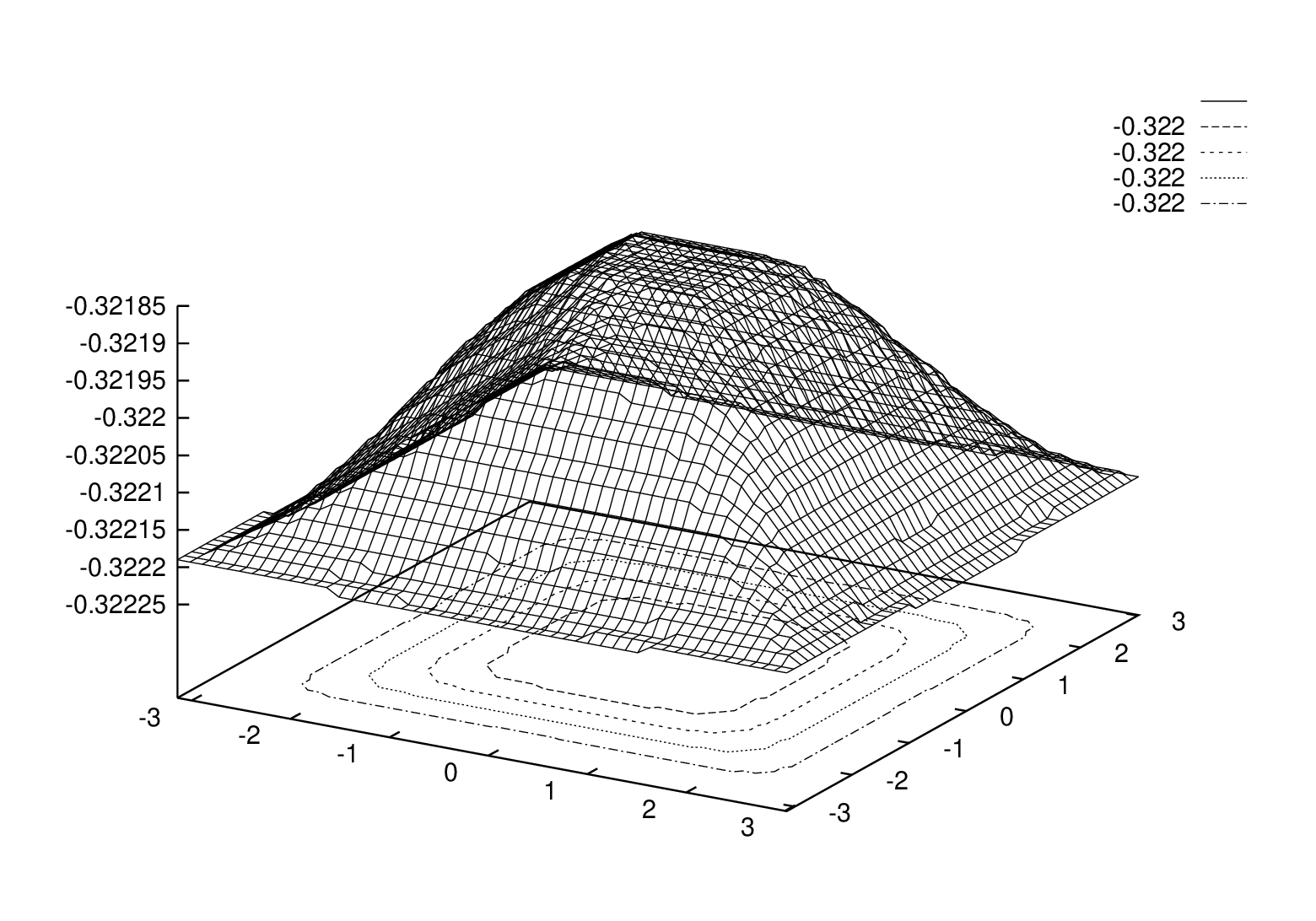}
\includegraphics[width=0.4\textwidth,height=0.31\textwidth]{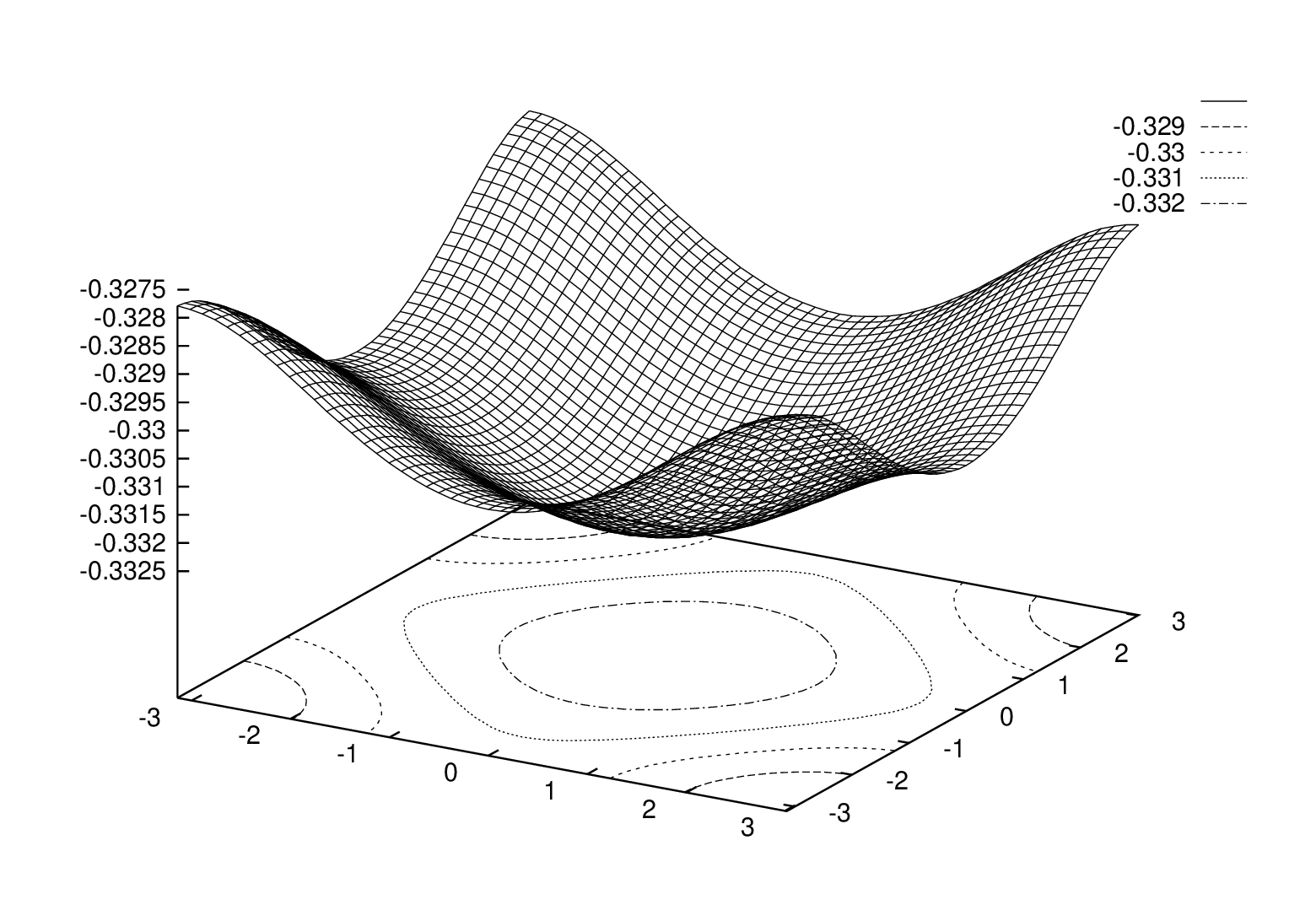}
\end{center}
\caption{Four bipolaron bands $E_{\nu}(K)$ versus wave-vector $K$ computed
close to
the triple point where bipolarons (S0), (S1) and (QS) are degenerate in energy
$(\alpha=0.017, U=0.232, t=0.08)$. Energies increase from bottom to
top.}
\label{4bandesG1}
\end{figure}

Because of the translation invariance of the model,
$M_{(S,i),(S^{\prime},j)}$ and
$P_{(S,i),(S^{\prime},j)}$ only depends on $j-i=n$. As a result,
eq.\ref{Sch1} can be diagonalized as combinations of plane waves with the form
$B_{S,j}(K)= \sum_{S} B_{S}(K) e^{i K j}$  with wave  vector $K$
which fulfills the eigenequation
\begin{equation}
	\textbf{P}^{-1/2}(K).\textbf{M}(K).\textbf{P}^{-1/2}(K).\textbf{B} =
	E_{\nu}(K).
	\textbf{B}(K)
	\label{Sch2}
\end{equation}
with the Fourier coefficients
\begin{eqnarray}
	\textbf{P}_{S,S^{\prime}}(K) & = &  \sum_{n}
P_{(S,j),(S^{\prime},j+n)} e^{i K n}
	\label{FourcoefP}  \\
	\textbf{M}_{S,S^{\prime}}(K) & = &  \sum_{n}
M_{(S,j),(S^{\prime},j+n)} e^{i K n}
	\label{FourcoefM}
\end{eqnarray}
Then the diagonalization of the $4 \times 4$ matrix\\
$\textbf{P}^{-1/2}(K).\textbf{M}(K).\textbf{P}^{-1/2}(K)$ yields the
eigenenergies\\ $E_{\nu}(K)$.
Figure (\ref{4bandesG1}) shows an example of calculation of these
four bands in the vicinity of the triple point.
Thus when there are four bipolarons that are
metastable (e.g. in the vicinity of the triple point of the phase
diagram) one obtains four bipolaron bands $E_{\nu}(K)$.
Within our approach the number of bipolaron bands is equal to
the number of metastable states for the adiabatic bipolaron which
provides the base about which we expand the eigenstates.
In other regions of the phase diagram the number of metastable
bipolarons changes, which induces (unrealistic) discontinuities for the
number of
bands. For example, when $U=0$ only the bipolaron (S0) is metastable, and
there is only one bipolaron band.

However the lowest bipolaron band does not exhibit very sharp
changes despite a small  discontinuous variation.
The reason that the upper bands are not reliable is that
the energies of these states might be also degenerate with phonon
excitations of the bipolaronic states of the lower band. The
real excited states
involve complex hybridization between these states.

Conversely, the bipolaronic states with the lowest energies should
not hybridize significantly with the higher energy states involving phonon
excitations.
Thus we consider that the lowest-energy
bipolaron band
provides a reliable description of the bipolaron excitations
close to its ground-state.
We use it to measure the bipolaron effective mass, that is,
the inverse of the curvature $T_{b}$ at zero wave
vector $K=0$. $T_{b}$ is constant in all directions because
of the square symmetry (fig.\ref{4bandesG1}). It can be viewed as
the effective hopping coefficient for the bipolaron tunnelling through
the lattice and can be compared with the prediction of \cite{ARR86}
given by formula (\ref{bipoltrint}), which is valid at both $U$ and $t$ small.

\begin{figure}
\begin{center}
\includegraphics{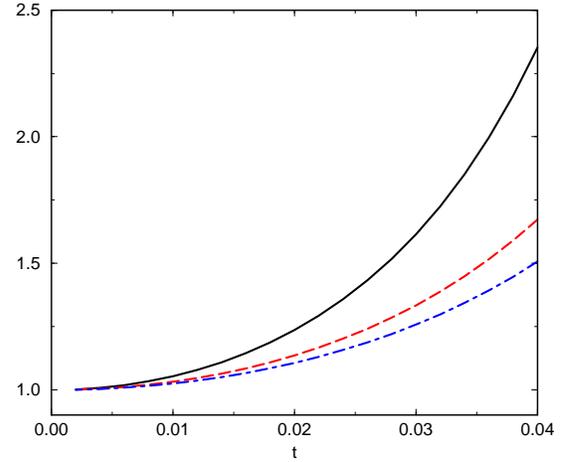}
\end{center}
\caption{Ratio $T_b/t_b$ versus $t$ of the transfer integral
numerically calculated and analytically predicted by formula
\ref{bipoltrint} \cite{ARR86} ($U=0$ and $\alpha=0.022$
(dot-dashed line), $\alpha=0.017$ (dashed line),
 $\alpha=0.01$ (full line).}
\label{AAR}
\end{figure}

Fig.\ref{AAR} exhibits the ratio $T_b/t_b$ as a function of the
effective transfer integral $t$ for $U=0$ for several values of the
quantum parameter $\alpha$. For $U$ small  this ratio goes to $1$ when $t$
goes to $0$,
which confirms the validity of formula (\ref{bipoltrint})
predicted by Alexandrov et al. \cite{ARR86} in that regime.
We also note that beyond this regime when the parameters $(U,t)$ are larger
than $0$,
$T_{b}$ becomes significantly larger than $t_{b}$, or equivalently the
bipolaron effective mass calculated numerically drops faster than
predicted by (\ref{bipoltrint}).

\begin{figure}
\begin{center}
\includegraphics[width=0.3\textwidth]{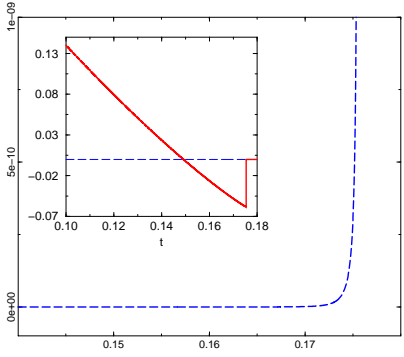}
\end{center}
\caption{Ratio $T_b/t$ versus $U$ at $t=0.04$ and $\alpha=0.01$; insert:
Bipolaron energy gain compared to a pair of free electrons versus $t$ at
$U=0$.}
\label{Effmbip}
\end{figure}

The insert of fig.\ref{Effmbip}
shows the bipolaron energy gain compared to a pair of free electrons
	\footnote{The energy gain compared to a pair of
		free electrons is not an accurate binding energy for the
		bipolaron.
		The bipolaron binding energy is precisely measured
		with respect to an unbound polaron state that is defined
		in the next section.}
at $U=0$ and the fig.\ref{Effmbip} shows the effective transfer integral
compared to the bare electronic transfer integral, which is  negligible at
the scale of the electronic energy. The bipolaron effective mass appears
to be much beyond than $10^{10}$ electronic masses even when its binding
energy vanishes. It is clear that this regime $U=0$ is not
favorable at all for the Bose condensation of such bipolarons that should
occur below a critical temperature inversely proportional to the effective
mass of the quasi-particle.

\begin{figure}
\begin{center}
\includegraphics[width=0.3\textwidth]{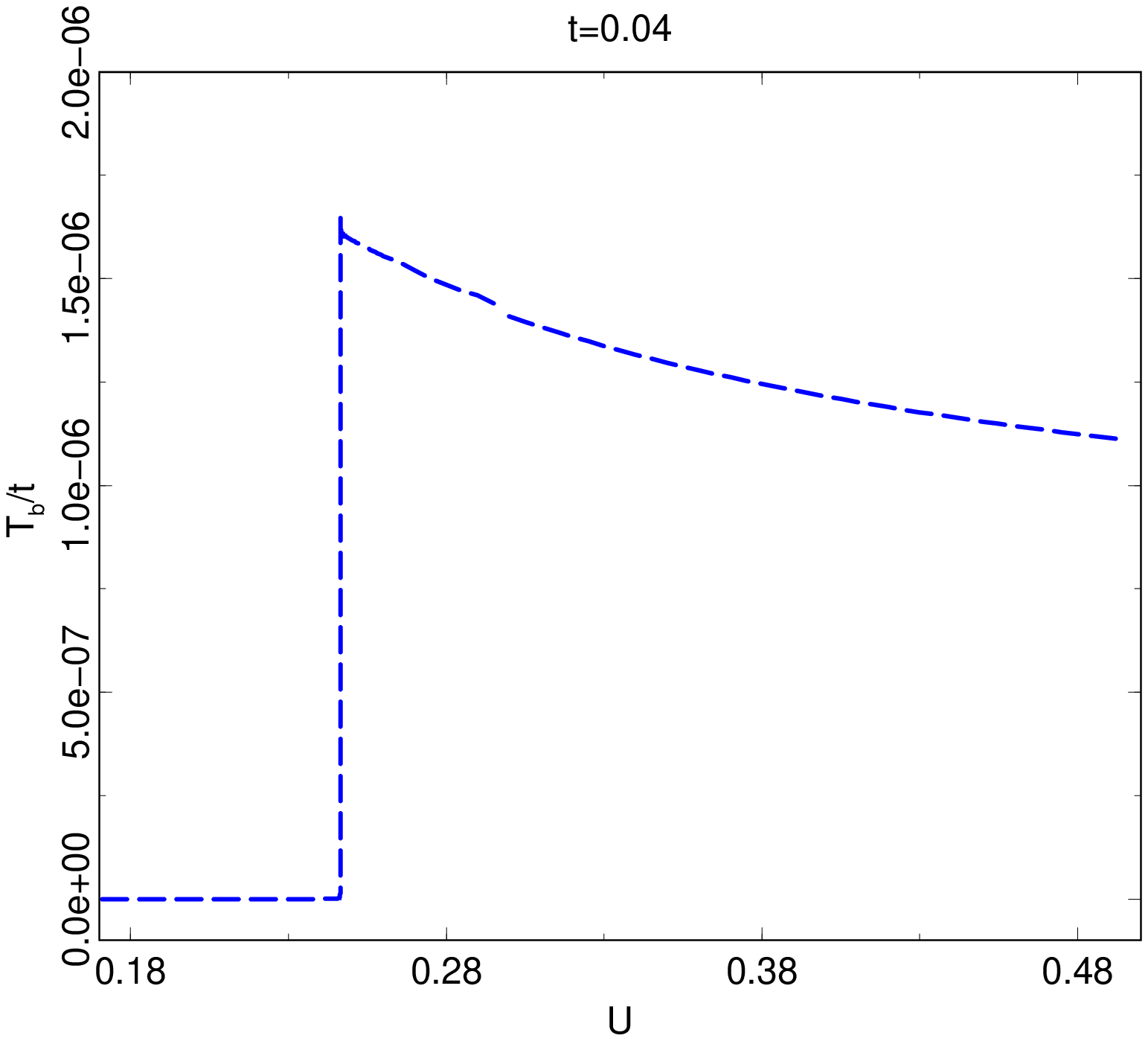}
\includegraphics[width=0.3\textwidth]{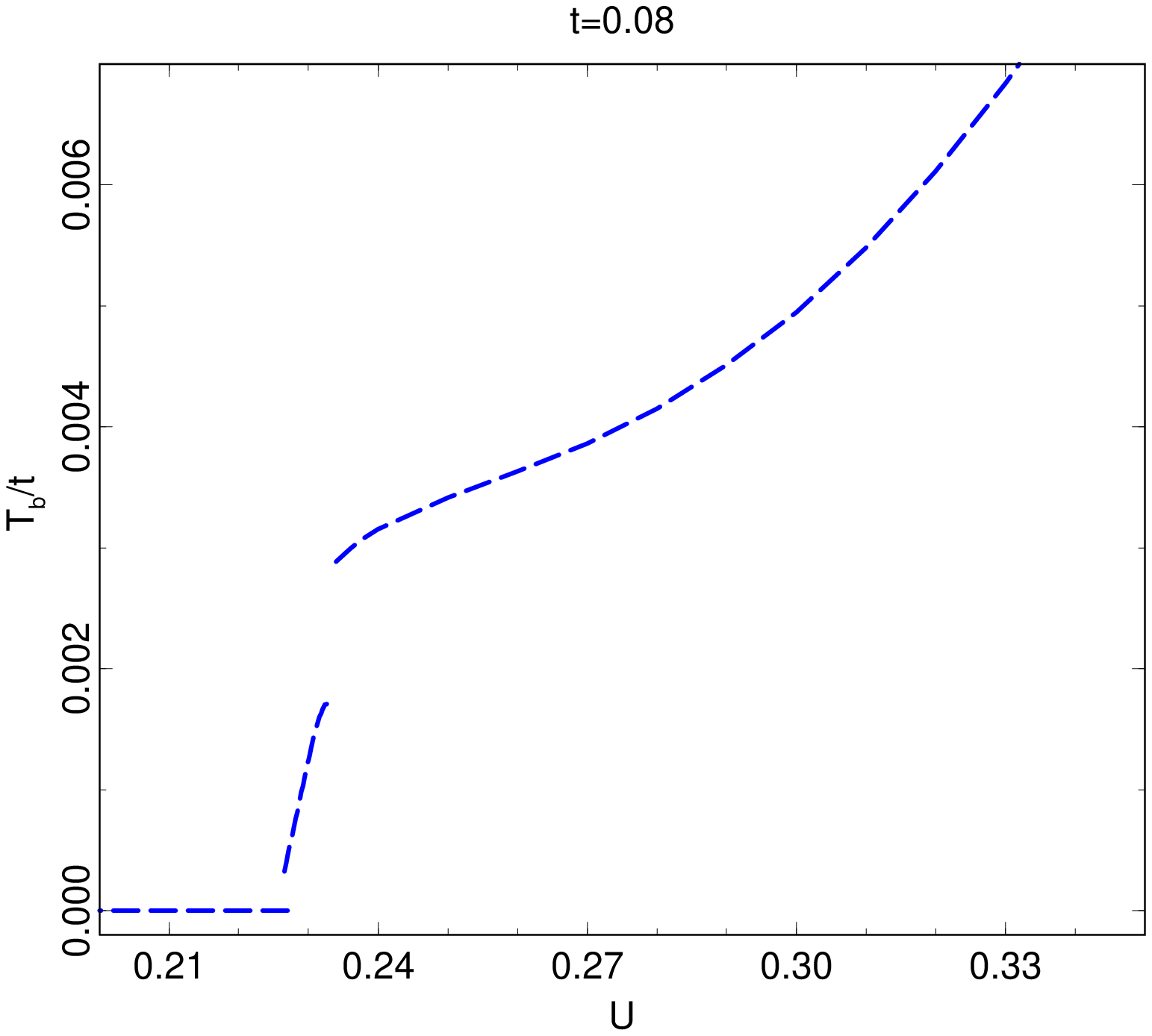}
\end{center}
\caption{Ratio $T_b/t$ versus $U$ at $t=0.04$ (top), $T_b/t$ versus $U$ at
$t=0.08$
(bottom) ($\alpha=0.01$).}
\label{Fig2b3}
\end{figure}

When the Hubbard term increases for relatively small
$t$, fig.\ref{Fig2b3} ($t=0.04$) shows that a sharp
discontinuity occurs when the
ground-state bipolaron becomes (S1) $U>0.25$.
There is a sharp increase of the
tunnelling energy by five orders of magnitude for this bipolaron
at $\alpha=0.01$.
In that case the Peierls-Nabarro barrier calculated in the previous
paper \cite{PA99} is still very high and consequently there is almost
no  hybridization between  (S0) and (S1). The smoothing of the
discontinuity of the tunnelling energy is thus hardly visible.

When $t$ is larger, the bipolaron (QS) becomes stable for
$U \approx 0.23$ and hybridization
between (S0), (S1) and (QS) becomes significant.
Actually the most important contribution to
the tunnelling energy of the bipolaron
comes from the hybridization
between (QS) and (S1). It is responsible
for the sharp increase of the tunnelling energy or
equivalently the sharp drop of the bipolaron effective mass.
This quantum mobility is favored when (QS) and (S1) are
degenerate in energy and separate by a weak Peierls-Nabarro barrier.
Then (QS) may tunnel to one of the four neighboring
(S1) and the latter tunnels to its neighboring (QS) that
corresponds to the initial one translated by one lattice spacing
in the direction of (S1) and so on.
The bipolaron tunnelling energy could reach $10^{-3}$ the bare
electronic energy which is not negligible anymore.

Let us point out that such a high mobility cannot be obtained within the
approximations used in \cite{ARR86} which do not consider
the possible degeneracies of several bipolarons.
The conclusions of \cite{CRF98} about the physical impossibility
of bipolaronic superconductivity are irrelevant for that situation.

% ***************************************
% ***************************************
\section{Variational Calculation of Quantum Polarons}
% ***************************************
\label{sect4}
In principle the above approach is valid only for very small
$\alpha$: that is, when the quantum lattice fluctuations are small.
However these fluctuations may increase drastically,
especially close to the first-order transitions when there are several
degenerate
bipolarons that we are especially interested in. Thus it is worthwhile to
improve our previous calculations by a variational approach
which should be
equivalent to the mean field perturbation  for small quantum lattice
fluctuations.

Our purpose is now to develop a quantum version of
the variational forms \cite{KAT98} used and
tested in the adiabatic case but which could hold for larger values of
$\alpha$.
Our approach is a simplified version of those of Toyozawa (see
\cite{Toy63} and \cite{ZBL97} for a recent application to the polaron in 1D).
We first test this method for the single polaron and will extend it in
the next section for the bipolarons of the Holstein-Hubbard model.

Because of the invariance of the system under translations the wave
function of a quantum single polaron is written as
a Bloch wave:
\begin{equation}
|\Omega^{P}(K)>= \frac{1}{\sqrt{\Lambda}} \sum_{n} e^{-iK.n}|\Psi^{P}(n)>
\label{ToPol}
\end{equation}
where $\Lambda$ is a normalization factor and $|\Psi^{P}(n)>$
is obtained from a unique wave function $|\Psi^{P}(0)>$ changing all
the indices $i$ of its electronic and atomic variables into $i+n$.
This transformation is nothing but a shift of the wave function from
site $0$ to $n$.

\subsection{Toyozawa approximation}
A simple variational approximation proposed by Toyozawa is to assume
that the local wave function is similar to the mean-field
polaron:
\begin{equation}
	|\Psi^{P}(j)> = \sum_{k} \left(\psi_{k-j}^P C_{k}^{+} \right).
	 \exp{\left(i \sum_l v_{l-j}. p_{l}\right)} |\emptyset>
	\label{elecgspol}
\end{equation}
To simplify the spin of the electron is omitted.
For each wave vector $K$ the variational energy
\begin{eqnarray}
<\Omega^{P}&(K)&|H|\Omega^{P}(K)> = \nonumber \\
&& \frac{\sum_{p} e^{i K p}<\Psi^{P}(j)|H | \Psi^{P}(j+p)>}{\sum_{p}
 e^{i K p}<\Psi^{P}(j)|\Psi^{P}(j+p)>}
	\label{vform}
\end{eqnarray}
is a function of the scalar products which does not depend on  $j$
  \begin{eqnarray}
	<\Psi^{P}(j)|\Psi^{P}(j+p)>&=&\exp - \frac{1}{4 \alpha} \sum_{i}
	(v_{i+p}-v_{i})^{2}
 \nonumber \\ &&        \times \left(\sum_{i} \psi_{i+p}^{*} \psi_{i}\right)
	\label{scpr2}  \\
	<\Psi^{P}(j)|H | \Psi^{P}(j+p)> &=&
	\exp - \frac{1}{4 \alpha} \sum_{i} (v_{i+p}-v_{i})^{2} \nonumber \\
	 \quad \times [( \sum_{n} \frac{1}{2} (\alpha+
	 v_{n+p} v_{n})) &\times& (\sum_{i} \psi_{i+p}^{*}
\psi_{i})\nonumber \\
 +\frac{1}{4} \sum_{n} (v_{n+p}+v_{n}) \psi_{n+p}^{*}
	\psi_{n} &-&\frac{t}{2} \sum_{i} \left(\psi_{i+p}^{*} \Delta
\psi_{i}\right)]
	\label{hamscp2}
 \end{eqnarray}
 and has to be extremalized with respect to the $2N$ parameters
 corresponding to the electronic wave function
$\{\psi_{j}^{P}\}$ and the lattice distortion $\{v_{l}\}$.
This form becomes exact in the adiabatic limit and should
improve the previous perturbation theory
as it is self-consistent.

A relation between the electronic density and the
average of the atomic displacement can be easily taken into account in
this variational form.
First, let us recall that the true eigenfunctions of $H$ are extrema of
$<\Psi|H|\Psi>$ in the full space of normalized functions $\Psi$.
For a given normalized eigenfunction $\Omega$ of $H$ we can
consider the one parameter family of normalized
functions $\Psi(\delta) = \exp{(i \delta . p_{j} )} |\Omega>$ where
the coordinate $u_{j}$ of the atom $j$ is changed into $u_{j}+\delta$.
The variational energy of this wave function is
$<\Psi(\delta)|H|\Psi(\delta)>$
which is equal to $<\Psi(0)|H_{\delta}|\Psi(0)>$
where
$H_{\delta}=\exp{-(i p_{j}\delta)} H \exp{(i p_{j} \delta)}$ is
simply obtained from $H$ by changing $u_{j}$ into $u_{j}-\delta$.
The variational energy
\begin{eqnarray}
&& <\Psi(\delta)|H|\Psi(\delta)> =<\Omega|H|\Omega> \nonumber \\
&-&\frac{1}{2}
\delta \left( 2<\Omega |u_{j}|\Omega> +<\Omega |n_{j}|\Omega>\right)
 + \frac{1}{2} \delta^{2}
	\label{vrfextr}
\end{eqnarray}
should be extremal for $\delta=0$, which implies
\begin{equation}
	<\Omega |u_{j}|\Omega> = - \frac{1}{2} <\Omega |n_{j}|\Omega>.
	\label{denpos}
\end{equation}
This result is nothing but an extension to the non-adiabatic case of the
standard relation  between the average atomic positions and the electronic
densities.

If we now consider an extremum of $<\Omega | H | \Omega>$ for $|\Omega>$
normalized
in the variational space defined by eq.(\ref{ToPol}) and (\ref{elecgspol})
this space is no longer invariant under the unitary operator $\exp \{i \delta
p_{j}\}$,
but it still remains globally invariant under operator  $\exp \{i \delta
\sum_{j} p_{j}\}$ which
performs a uniform displacement by $\delta$ of all the atoms. We
apply the same argument  as above that is, study
$<\Omega(\delta) | H | \Omega(\delta)>$ where $\Omega(\delta)=\exp \{i
\delta \sum_{j} p_{j}\}\Omega$
is extremal for $\delta=0$. This condition yields
$\sum_{j} <\Omega |u_{j}|\Omega> = - 1/2 \sum_{j} <\Omega |n_{j}|\Omega>$.
For the variational
extrema with the Toyozawa form (\ref{ToPol}) and (\ref{elecgspol}),
we find $<\Psi(l)|\sum_{n}u_{n}|\Psi(m)> = (\sum_{n} v_{n})
<\Psi(l)|\Psi(m)>$,
which readily implies $\sum_{j} <\Omega |u_{j}|\Omega> = \sum_{n} v_{n}$.
For the polaron, that is, for a system with only one electron the extremum
of the Toyozawa form (\ref{ToPol}) and (\ref{elecgspol})
necessarily fulfills
\begin{equation}
    \sum_{n} v_{n}= - \frac{1}{2}
	\label{denposf}
\end{equation}

\subsection{Toyozawa Exponential Approximation: TEA}

Minimizing the variational form (\ref{vform}) for the whole set of $2N-1$
parameters $\{\psi_{i}\}$
and $\{v_{i}\}$ with condition (\ref{denposf}) is a complex numerical task
which moreover will become
even more complex when extended to the bipolaron problem. However, we can
expect that the behavior
of the variational parameters  $\{\phi_{n}\}$ and $\{v_{n}\}$
will not be far from exponential at infinity.
Thus assuming simple exponentials for $\{\phi_{n}\}$ and $\{v_{n}\}$
should not be a bad approximation
 as proposed in \cite{KAT98} at the adiabatic limit.
 Taking into account the normalization and condition (\ref{denposf})
  we postulate that the electronic wave function and the atomic
modulation have the form:
\begin{eqnarray}
\psi_{i}^P&=&A\lambda^{|i_x|+|i_y|} \quad  A^{-1}=(1+\lambda^2)/(1-\lambda^2)
\label{TEApolElec} \\
v_{i}^P&=& -B \mu^{|i_x|+|i_y|}  \quad   B^{-1}=2(1+\mu)/(1-\mu)
\label{TEApolPhon}
\end{eqnarray}

for each wave vector $K$ there are only two variational parameters
$\lambda(K)$ and $\mu(K)$ instead of $2N$ for the original  Toyozawa ansatz
which  allows much simpler calculations although we still need a
numerical minimization of (\ref{vform}). To that aim
we use a simplex method \cite{NumRec}, which
 is the most efficient algorithm we tested because it avoids
any precision problem due to the numerical computation of the derivatives.

The Toyozawa Exponential Ansatz (TEA) turns out to be almost as good
as the full ansatz when the polarons are small, since in that case the
exponential approximates quite well its shape.
When the size of the polaron becomes larger, the TEA (as well as the
original Toyozawa ansatz) yields a first order transition. This first order
transition
is well-known to exist at the adiabatic limit $\alpha=0$ where the ground
state of a single electron
undergoes a first order transition from a small polaron to a free electron
\cite{HE76,KAT98} at $t=t_{p}\approx 0.07486$.

\begin{figure}[tbp]
\begin{center}
\includegraphics{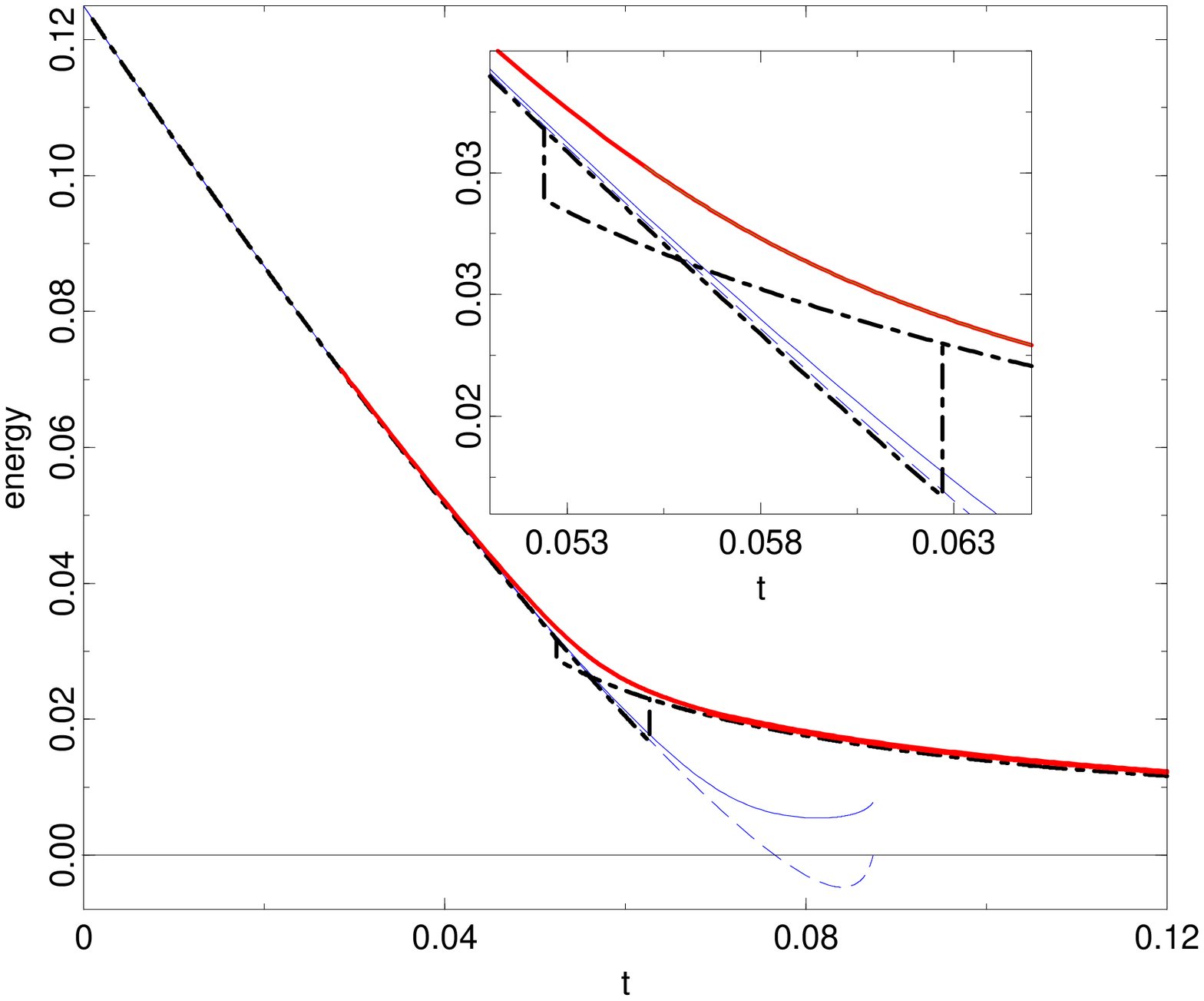}
\caption{Binding energy of the quantum polaron versus $t$ at $\alpha=0.017$
calculated with several approximations as explained in the text
and magnification (insert).}
\label{ElPolG3}
\end{center}
\begin{center}
\includegraphics{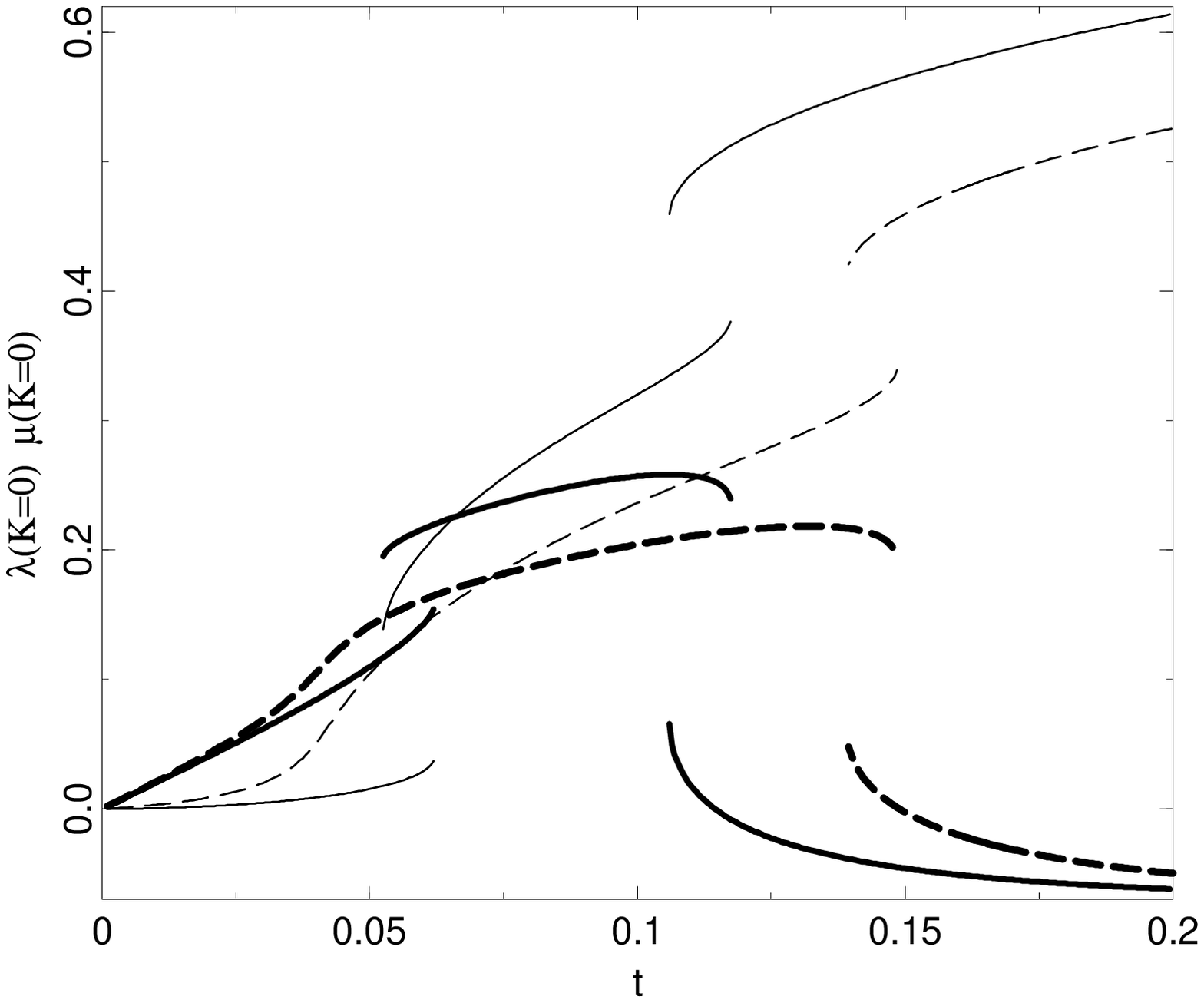}
\caption{Variational parameters versus $t$ for the TEA approximation of the
polaron
$\lambda(K)$ (thick lines) and $\mu(K)$ (thin lines).
$\mu(K)$ (thin lines). Wave vector $K$ is zero and
$\alpha=0.017$ (full lines) and $\alpha=0.03$ (dashed lines)}
\label{Min}
\end{center}
\end{figure}

We define the binding energy of the quantum polaron
as the difference between the energy of the
extended electron at zero wave vector $K=0$ and that of the bottom
of the polaron band. Fig.\ref {ElPolG3} shows the
variation of the binding energy
versus $t$ for the quantum polaron calculated in
several different approximations including the assumption that:
  \begin{enumerate}
	\item  The polaron band is calculated as for the bipolaron bands
	(section \ref{BBW}) from perturbation of the mean field polaron
	(thin dashed line) \label{approx1};

	\item  The polaron band is hybridized with the free electron band
	(thin full line) \label{approx2};

	\item The polaron band is calculated with the TEA approximation
  (thick dashed line) \label{approx3}; and

	\item  The polaron band is calculated with the HTEA approximation
	where small and large polarons are hybridized
	(thick full line) (see next section for details) \label{approx4}.
  \end{enumerate}

When the quantum lattice fluctuations are small (which occurs either at the
adiabatic
limit $\alpha=0$ or when $t$ is small), these approximations yield
practically the
same result. When $\alpha \neq 0$ the best variational form is that
which gives the lowest energy for
the ground-state (that is, the largest binding energy).
The results of
these approximations become significantly different when $t$ approaches
the critical value $t_{p}$ at which the adiabatic first order transition
occurs. Each of these approximations improves the previous one,
since the polaron energy becomes lower at each step.

It is clear that approximations (\ref{approx1}) and (\ref{approx2}),
which keep the polaron shape
rigidly fixed to that at the adiabatic limit, are not appropriate
to remove the first-order transition (see \ref{ElPolG3}).
The TEA approximation (\ref{approx3}) also yields first-order transitions,
but there are two distinct transitions occurring at $t=t_p^1(\alpha)<t_p$
and $t=t_p^2(\alpha)>t_p$
and the amplitudes of energy discontinuities are weak
because the polaron shape is determined self consistently.

Fig.\ref{Min} shows $\lambda$ and $\mu$ values that minimize (\ref{vform})
at $K=0$. The first of the TEA transitions ($t_p^1(\alpha)<t_p$)
occurs between a small and a large polaron
(see fig.\ref{Min}) and $t_p^1(\alpha)$ decreases when $\alpha$ increases
before that transition disappears
for $\alpha>0.03$.
The second TEA transition ($t_p^2(\alpha)>t_p$)
persists for large ($\alpha>0.7$)
but it is hardly distinguishable on the binding energy plot
(fig.\ref{ElPolG1}).
The transition occurs between a large polaron and a quasi-free electron
with an extended phonon part (that is, $\mu(K=0)$
tends to $1$ when $t$ is large).
Note also that $\lambda(K=0)$  may become negative
in the regime of large $t$ and small $\alpha$
but then the polaron binding energy becomes  negligible
so that it is meaningless to use a polaron picture
for a regime that is better described as a Fermi liquid.

\subsection{Smoothing the First Order transition: HTEA}

Actually, any first-order transition  for the polaron ground-state
(or the bipolaron) which would be obtained by any variational method
cannot exist physically.
The reason is that at the transition point there are two
approximate wavefunctions with the same variational energy which are
supposed to approximate the ground-state. It is
possible to hybridize these two
degenerate states to obtain a new state with a lower energy.
The same arguments hold for the exact ground-state, which cannot
exhibit any first-order transition.

On the basis of these arguments, we demonstrate numerically
that the two first-order transitions obtained with the TEA of polaron
can be smoothed using a variational form for the wave function
$\Psi^{P}(0)$ in eq.\ref{ToPol} which  hybridizes three wave functions,
\begin{equation}
\Psi^{P}(0)=\beta_{1} \Psi_{1}^{P}(0) + \beta_2 \Psi_{2}^{P}(0)+ \beta_3
\Psi_{3}^{P}(0).
	\label{hybwf}
\end{equation}

Each of these wave functions has the TEA form (\ref{TEApolElec}) and
(\ref{TEApolPhon})
with parameters $\lambda_{1},\mu_{1}$, $\lambda_{2},\mu_{2}$ and
$\lambda_{3},\mu_{3}$ respectively.
Hybridizing three wave functions instead of two has the advantage of
sweeping out simultaneously the two successive first order transitions.
The variational energy (\ref{vform}) now depends on $9$
parameters $\lambda_{S},\mu_{S},\beta_{S}$ with $S\in \{1,2,3\}$.
Let us note
\begin{eqnarray}
M_{S,S'}(K)&=&\sum_{p} e^{i K p}<\Psi^{P}_{S}(j)|H |
\Psi^{P}_{S'}(j+p)>\nonumber \\
	\label{Ediag1}
\end{eqnarray}
and
\begin{eqnarray}
P_{S,S'}(K)&=&\sum_{p} e^{i K p}<\Psi^{P}_{S}(j)|
\Psi^{P}_{S'}(j+p)>\nonumber \\
	\label{Ndiag1}
\end{eqnarray}
where $(S,S') \in \{1,2,3\}^2$.
We point out that because of the central symmetry of the TEA,
the $3 \times 3$ matrices $M$ and $P$ are real.
Then the energy of the ground-state $E(K)$ has the
following variational form:
\begin{eqnarray}
<\Omega^{P}(K)|H|\Omega^{P}(K)>=
\frac{\sum_{S,S'} \beta_{S}\beta_{S'}^{*} M_{S,S'}}
{\sum_{S,S'} \beta_{S}\beta_{S'}^{*} P_{S,S'}}
	\label{Ediag2}
\end{eqnarray}

The extremalization of $E(K)$ eq.\ref{Ediag2} with respect to $\beta_{1}^{*}$
$\beta_{2}^{*}$ and $\beta_{3}^{*}$ yields the set of three equations
\begin{eqnarray}
\sum_{S} \beta_{S} M_{S,S'}- E(K) (\sum_{S} \beta_{S} P_{S,S'})=
0 , \label{Ediag3}
\nonumber
\end{eqnarray}
and therefore we have to solve eigenvalue problem
$M {\bf \beta}= E(K) P {\bf\beta}$:  that is,
$E(K)$ is the lowest eigenvalue of the matrix $P^{-1/2}(K) M(K) P^{-1/2}$.
That calculation is very similar to the perturbative method of the mean-field
described previously in the case of the bipolaron
but here the lowest eigenvalue $E(K)$
has still to be minimized with respect to the set of six parameters
$(\lambda_{S},\mu_{S})$.

\begin{figure}[tbp]
\begin{center}
\includegraphics{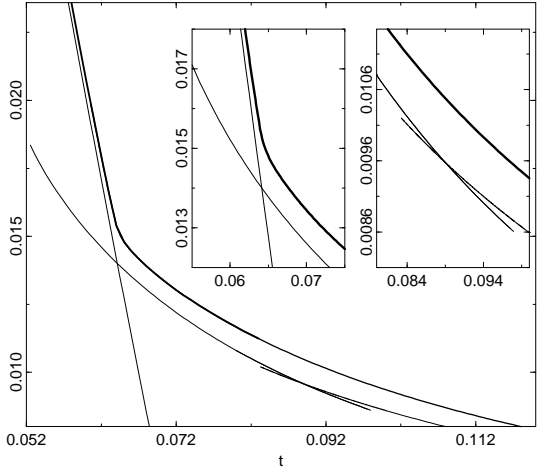}
\caption{Binding energy of the quantum polaron versus $t$
calculated with TEA (thin lines) and the HTEA (thick lines) ( $\alpha=0.01)$.
Magnification of the two first order transitions of the TEA (inserts).}
\label{ElPolG1}
\end{center}
\end{figure}
For small $t$ we recover the TEA results
(that is, only one $\beta$ is nonnegligible, fig.\ref{ElPolG3}).
Close to the TEA first-order transitions
the variational ground-state appears as the hybridization of
either a small polaron and a large polaron or a large polaron and a
quasi-free electron (very large polaron).
A significant increase of the binding energy of polaron results from this
hybridization in these crossover regions where
the first order transitions are smoothed and thus removed fig.\ref{ElPolG1}.
Furthermore our calculations show that the energy gain due to hybridization
persists for large $t$ values. In that regime
the fluctuations of the quantum lattice are
strong enough to hybridize two TEA states, the
large polaron and the quasi-free electron, whose energies
differ only slightly.

\begin{figure}[tbp]
\begin{center}
\includegraphics{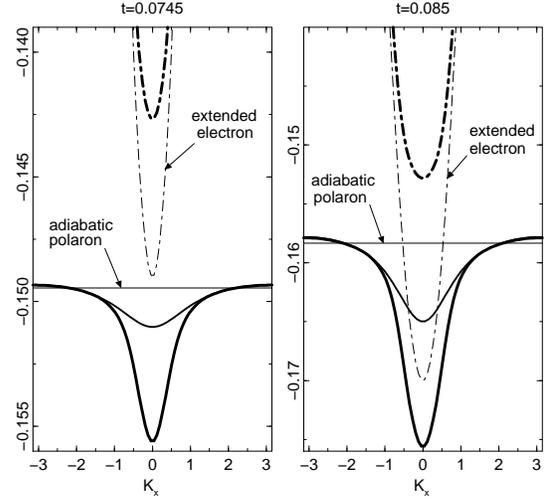}
\end{center}
\caption{Polaron Bands $E_{S}(K)$ and free electron band $E_{L}(K)$ in the
$K_{x}$
direction at $\alpha=0$ (thin lines) and $\alpha=0.017$ calculated 1) as a
perturbation
of  the adiabatic polaron (thick lines) and 2) by hybridizing the polaron
band and the free electron
band (thicker lines)
at $t=0.0745<t_p$ (left) and $t=0.085>t_p$ (right).}
\label{bandesPol}
\end{figure}

\begin{figure}[tbp]
\begin{center}
\includegraphics{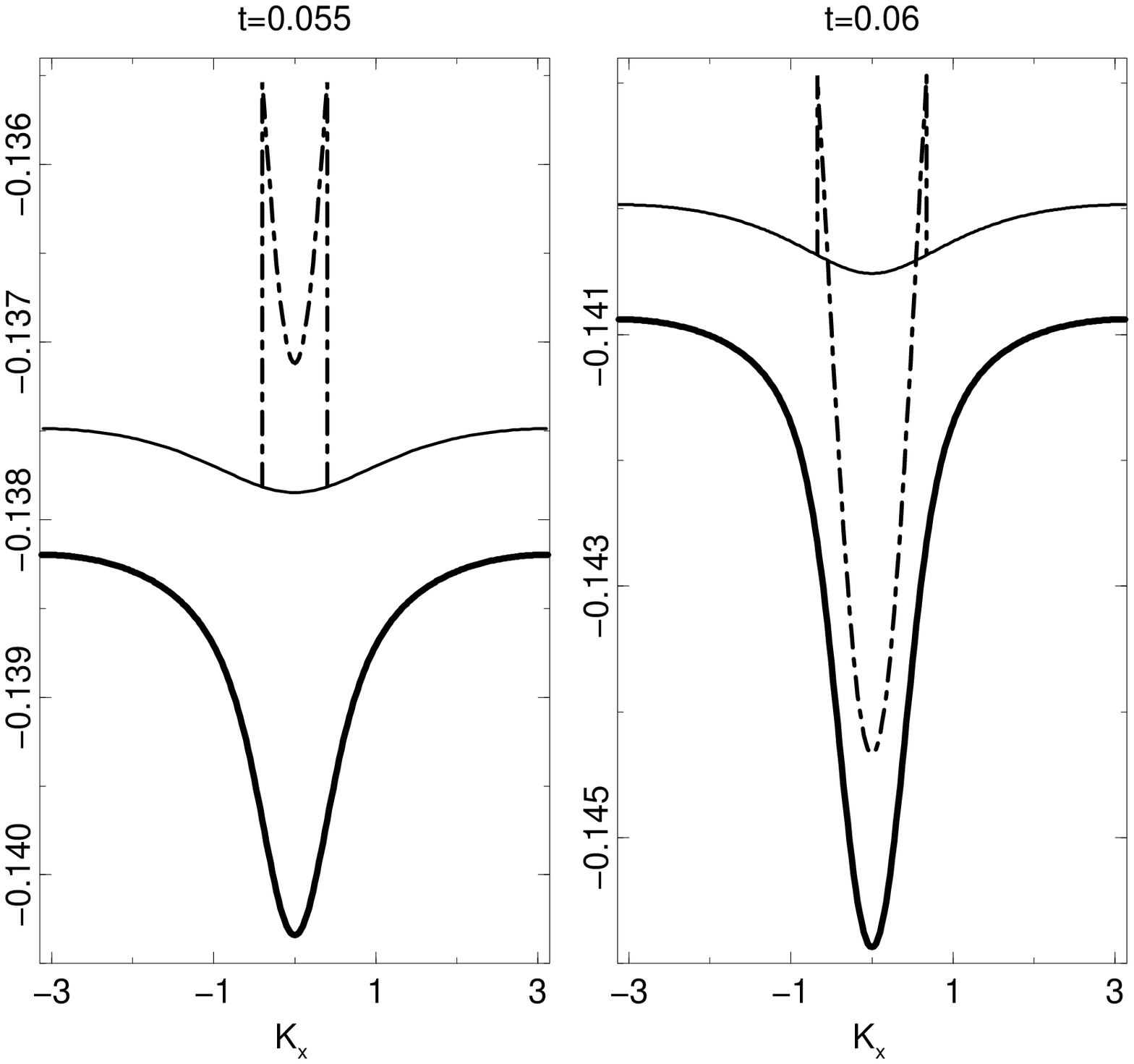}
\end{center}
\caption{TEA Bands (thin lines) and HTEA Bands (thick lines)
$E_{S}(K)$ (full lines) and $E_{L}(K)$ (dot-dashed)
in the $K_{x}$ direction at $\alpha=0.017$
at $t=0.055<t_p^1(\alpha)$ (left) and $t=0.06>t_p^1(\alpha)$ (right)}
\label{fig6b}
\end{figure}

A consequence of the hybridization can be also observed on the shapes
of the polaron bands. In the adiabatic limit ($\alpha=0$), the small polaron
is degenerate
under arbitrary lattice translations, which means that the
polaron band is perfectly flat, as shown on fig.\ref{bandesPol}. In the
regime where the polaron is metastable for $t>t_{p}$, the flat polaron band
intersects the free electron band so that there is a line of wave vectors
where the small polaron state and the free electron state are degenerate
(see fig.\ref{bandesPol}).
With nonvanishing quantum lattice fluctuations ( $\alpha \neq 0$),
the degeneracies are lifted along the intersection line.
Approximation \ref{approx2} provides an important correction in the vicinity
of $t_p$ where the adiabatic polaron becomes extended.
Around the degenerate line there is a cross-over region in wave vector
where the component of the free electron to the ground-state
varies from almost 1 to almost 0 when $K$ goes from
$0$ to $\pi$ and the opposite for the component of the small polaron.
Thus there is a smooth exchange of the quantum state
from the large to the small polaron  (fig.\ref{fig6b}).
This exchange also occurs for the upper band but in reverse order from
the small polaron to the large polaron.

In the band of the TEA, for $t^1(\alpha)<t<t^2(\alpha)$
only one  first-order transition is observed   in $K$ space
between a small polaron and a large one.
For $t>t^2(\alpha)$
two first order transition might be observed
in $K$ space at different $K$ values and
they occur between the three kinds of polaron describded previously.
They are smoothed with the HTEA.

\begin{figure}
\begin{center}
\includegraphics{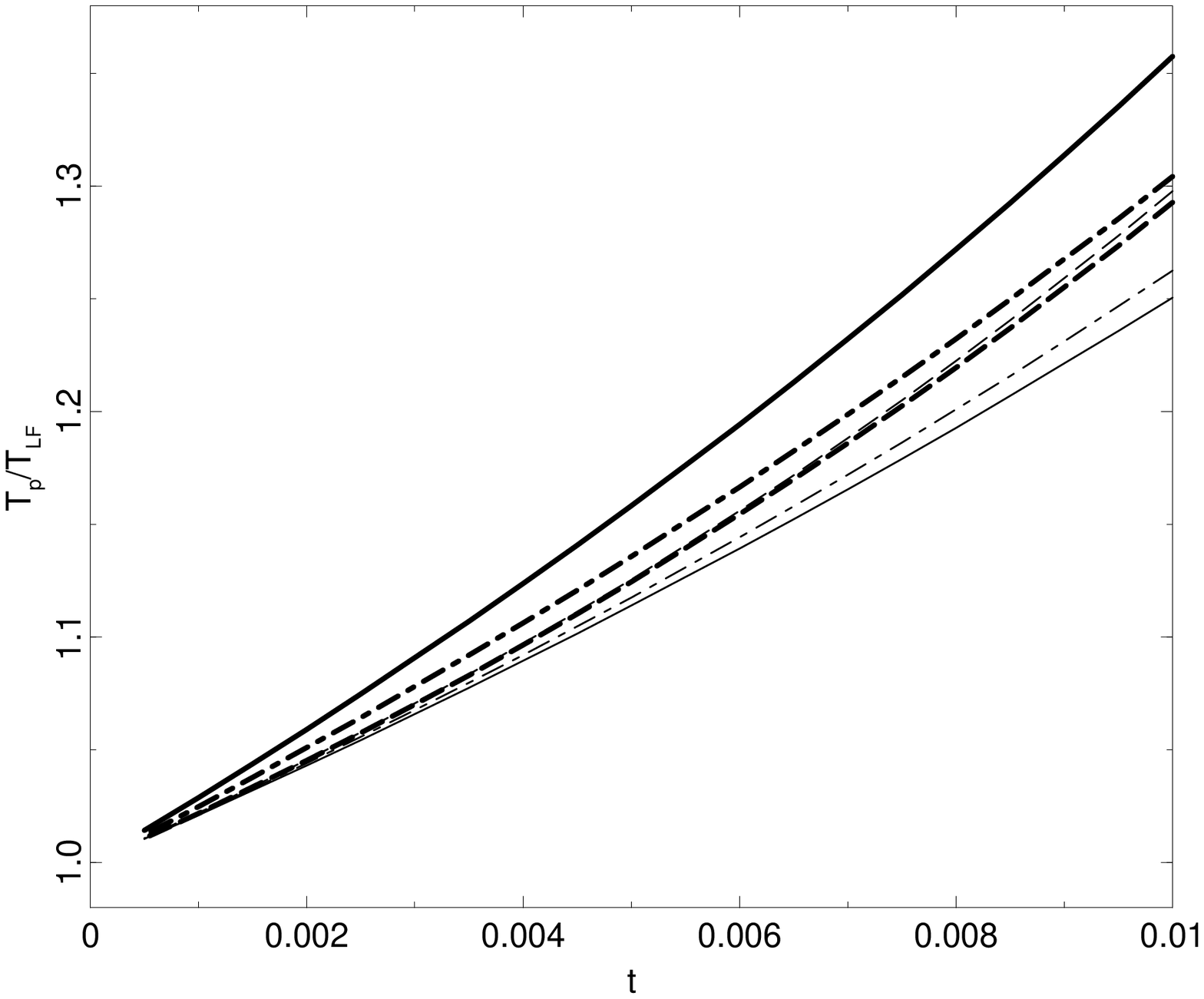}
\caption{ Ratio $T_{p}/T_{LF}$
versus $t$  at $\alpha=0.01$
(dashed line), $\alpha=0.022$ (dot-dashed lines), $\alpha=0.03$ (full line)
calculated as a perturbation to the adiabatic limit (thin lines)
and using the  TEA (thick lines).}\label{fig6c}
\end{center}\end{figure}

\begin{figure}
\begin{center}
\includegraphics{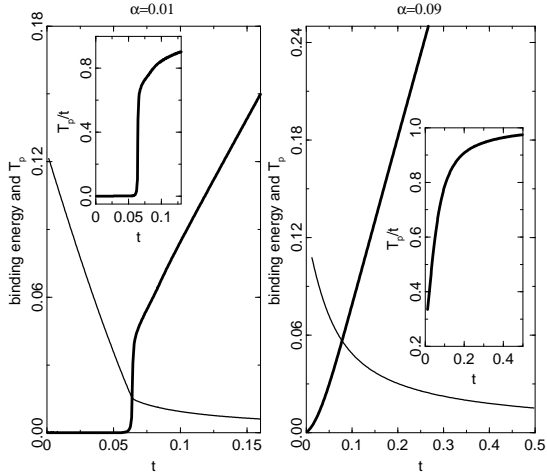}
\caption{ Polaron tunnelling energy  $T_{p}$ (thick lines)
and its binding energy (thin lines) versus $t$  at $\alpha=0.01$ (left),
$\alpha=0.09$ (right) calculated with the HTEA method.
Inserts: Ratio $T_p/t$ versus $t$.
}\label{mHTEA}
\end{center}\end{figure}

The effective polaron mass is the inverse curvature $1/T_p$ at the bottom
of the lowest polaron
band at $K=0$. It can be calculated as a function of $t$ and $\alpha$
and compared
with the value $1/T_{LF}$ obtained from the Lang-Firsov transformation
(\ref{poltrint}).
The variation versus $t$ of the ratio $T_p/T_{LF}$ is shown
fig.\ref{fig6c} for different $\alpha$.
For $\alpha$ small, this ratio $T_{p}/T_{LF}$
is almost one, which confirms that
the mean-field approximation used
to establish formula (\ref{poltrint}) is valid for both
 quantum lattice fluctuations and $t$ small.
When $t$ increases from zero the ratio $T_{p}/T_{LF}$ increases
from unity, which means that eq.\ref{poltrint} overestimates the polaron
effective mass. We already observed this effect for the bipolaron case in
the formula eq.\ref{bipoltrint}. However this effect does not imply sharp
variations.

To compare the polaron mass and the bare electronic mass,
the ratio $T_p/t$ is plotted fig.\ref{mHTEA} for different values of $\alpha$.
For large $t$ as well as for large $\alpha$, the polaron effective mass
reduces to the bare electron mass. In other words the electron becomes
practically free.

When $\alpha$ is small, there is a sharp drop in  the
inverse effective polaron mass, which is
reminiscent of the first order transition at $\alpha=0$ between the localized
small polaron and the extended electron.
This sharp variation becomes  smoother and smoother as
$\alpha$ increases.

Fig.\ref{mHTEA} compares the binding energy of the polaron
and its  tunnelling energy $T_p$.
For small $t$, the binding energy is much larger than $T_p$,
while for large $t$ it becomes much smaller.
There is a value $t=t_c(\alpha)$
where  both energies are equal.  In the vicinity of this region
the polaron has maximum  mobility while it remains
reasonably well-bound (compared to this tunnelling energy!).

% ***************************************
% LES (TEA) du BIPOLARONS
% ***************************************
\section{Variational Calculation of Quantum Bipolarons}
\label{sect5}

The variational methods  (\ref{elecgspol}) we used for the single
polaron can be extended to bipolarons with variational forms  (S0), (S1)
and (QS).
For this purpose we write the bipolaron wave function as a Bloch wave:
\begin{equation}
|\Omega^{B}(K)>=\frac{1}{\sqrt{\Lambda}}\sum_{j} e^{-iK.j}|\Psi^{B}(j)>,
\label{ToPol3}
\end{equation}
and we postulate an extended Toyozawa form for the local wave function
\begin{eqnarray}
	|\Psi^{B}(0)> &=& \left(\sum_{j,k} \psi_{j,k}^{B} C_{j,\uparrow}^{+}
	C_{k,\uparrow}^{+}\right)
	\nonumber \\
	&\times & \exp {\left(i \sum_{l} v_{l}^{B} p_{l}\right)} |\emptyset>
	\label{Toybip}
\end{eqnarray}

\subsection{TEA quantum Bipolarons}
The simple TEA approximation for the bipolaron consists in  choosing
$\psi_{j,k}^{B}$ with the form (\ref{AS0}) for B=(S0), (\ref{AS1}) for
B=(S1) or (\ref{ASQ}) for B=(QS) and  $v_{l}$ with exponential forms which
depend on the type of bipolaron as follows:

\begin{eqnarray}
	v_{l}^{S0}(0)&=& -C_{S0} \mu_{S0}^{|l_x|+|l_y|}\label{depS0}\\
	v_{l}^{S1_x}(0)&=& -C_{S1}[ \mu_{S1}^{|l_x|+|l_y|}
+ \mu^{|l_x-1|+|l_y|}]\label{depS1}\\
	v_{l}^{QS}(0)&=& -C_{QS} \mu_{QS}^{|l_x|+|l_y|}.\label{depQS}\\
\nonumber
\end{eqnarray}

The same arguments used to prove equation (\ref{denposf}) imply
\begin{equation}
    \sum_{n} v_{n}^{B}= - 1,
	\label{denbisf}
\end{equation}
which determines the parameters $C_{S0}$, $C_{S1}$ and $C_{QS}$.
Using the scalar product formula (\ref{scpr}) and (\ref{hamscp}), the
variational
energy (\ref{vform}) is calculated numerically and minimized
with respect to both $\lambda$ and $\mu$ parameters for
each value of the wave vector $K$ and for each bipolaron (S0),(S1) or (QS)
(see fig.\ref{fig9b}).
This variational form still has  a small number of parameters which
allows a fast numerical minimization.

\begin{figure}
\begin{center}
\includegraphics{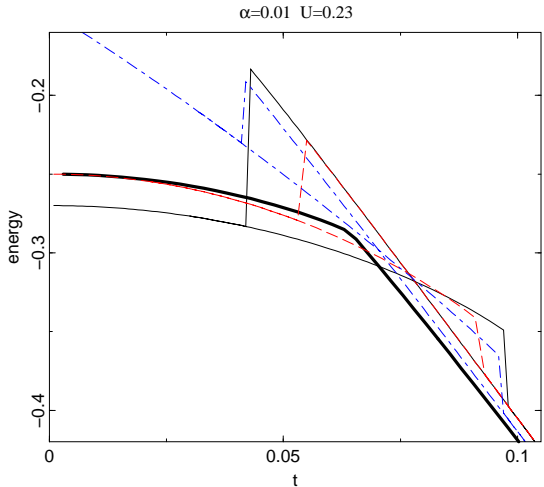}
\caption{TEA energy versus $t$ for bipolaron (S0) (thin full line), (S1)
(thin dashed line), (QS) (thin dot-dashed line)
and  energy of two single HTEA polarons (full thick line)
at $U=0.23$ and $\alpha=0.01$.}
\label{fig9p}
\end{center}\end{figure}

\begin{figure}
\begin{center}
\includegraphics{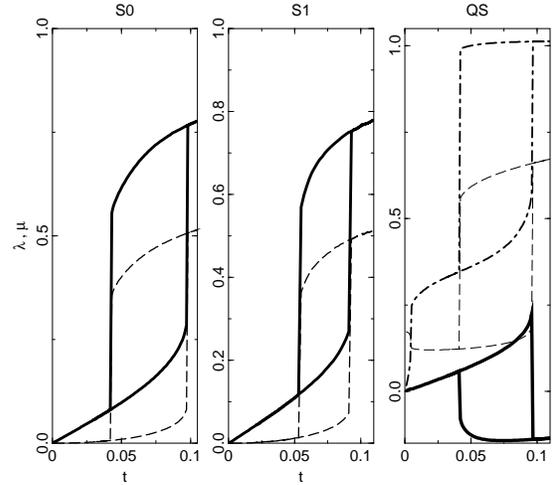}
\caption{Parameters $\lambda$ (full lines)  and $\mu$ (dashed lines)
versus $t$ for the minima of the variational
form (\ref{vform}) for the TEA bipolarons (S0),(S1) and (QS)
(from left to right) at $U=0.23$, $\alpha=0.01$ and the wave vector $K=0$. 
Note that (QS) (right figure) has two $\lambda$ variables
$\lambda_1$ (full line) and $\lambda_2$ (dot-dashed).}
\label{fig9b}
\end{center}\end{figure}

The minimum energy is always found to be at the bottom of the lowest band
at $K=0$. The quantum corrections to the energies of bipolarons
(S0), (S1) and (QS) are compared
with the energy of two  polarons far apart. We use the HTEA result
described in the previous section (fig.\ref{fig9p}), since we know that it
yields the
lowest and thus the most accurate energy for the quantum polaronic
ground-state.

As for the TEA polaron each TEA  bipolarons (S0), (S1) or (QS)  exhibits a
first-order transition
when $t$ increases between a small and a large bipolaron with the same
symmetry.
Actually if one compares the energies of all the possible solutions
these large
bipolarons are found never to be the ground-state whatever
$\alpha$ is, because
a pair of single quantum polarons has always less energy.
As a result, these bipolarons always gain energy by breaking up into two
polarons
(fig.\ref{fig9p}) even for large $\alpha$.

\begin{figure}
\begin{center}
\includegraphics{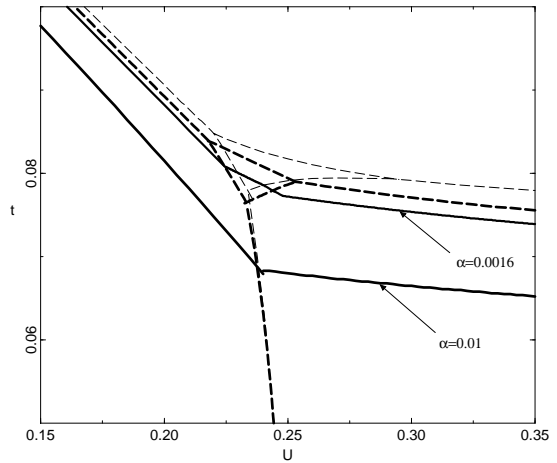}
\caption{Phase diagram of the TEA bipolarons (S0),(S1), (QS) and a pair of
unbound polarons for $\alpha=0$ (thin dashed line),
$\alpha=0.001$ (thick dashed line), $\alpha=0.0016$ (full thin line),
and $\alpha=0.01$  (full thick line).
The case $\alpha=0$ is the adiabatic case already calculated
in ref.\cite{PA99}.}
\label{fig9}
\end{center}\end{figure}

In the adiabatic limit ($\alpha=0$), these TEA calculations
become identical to the variational calculation which was described
in \cite{PA99} (see fig.\ref{fig9}).
Comparing the energies of these TEA bipolarons (without any hybridization)
we construct a new phase diagram for $\alpha$ non zero with first-order
transition lines
and test how it changes when the quantum lattice parameter $\alpha$ increases.

The approximate calculations of the tunnelling energy
for the polaron (\ref{poltrint}) and for the bipolaron (\ref{bipoltrint})
suggests that for $U$ sufficiently different from $ 1/4$ the tunnelling of a
single polaron with one electron
is much easier than that of a bipolaron, which contains two electrons and
moreover involves a bigger lattice distortion.
Thus, one should expect more generally that the energy gain
generated by the  quantum lattice fluctuations for the single polaron
is systematically much larger than that for the bipolarons.
As a result, the domain of parameters where the ground-state consists of an
unbound
pair of large polarons should extend at the expense of the domains of the
bipolarons when $\alpha$ increases.

Indeed fig.\ref{fig9} confirms that the first-order transition lines which
exist in the adiabatic limit shift to lower values of $t$ when $\alpha$
increases. As a consequence the domain of existence of the (QS)
bipolaronic
ground-state shrinks to zero for a rather small value (approximately $0.002$)
of $\alpha$ and completely disappears for larger values.

The disappearance of the triple point of the phase diagram between bipolarons
(S0),(S1) and (QS) for relatively small values of $\alpha$
seems to rule out our suggestion
that bipolarons could become
very light.
We show how to recover this possibility
in the last section by
minor changes in the model which may restore
this triple point for relatively large values of $\alpha$.

\subsection{HTEA quantum Bipolarons}

We carefully examined whether the HTEA calculation of bipolarons, which in
principle should be more accurate,  could change this conclusion.
Actually, it will not change it, and to not confuse or bother the reader,
all details of our unsuccessful (but useful)
numerical investigations are not presented.

As said previously, in principle no first-order
transitions could exist for the ground-state of a pair of electrons
interacting with the lattice. They are removed by
hybridization of all (or only those which are relevant), degenerate
bipolaron solutions
(S0),(S1) or (QS) both small and large, which necessarily generates some
energy gain.

The HTEA calculation for the bipolaron is similar to that for the polaron
except that it
may involve more bipolaronic states.
We assume generally that the wave function
$\Psi^{B}(0)$ (\ref{ToPol3})
is a normalized combination of $n$ wave functions ($n$ depends
on the number of TEA which hybridize)
\begin{equation}
\Psi^{B}(0)>=\sum_{S} \beta_{S} \Psi^{B}_{S}(0) >
	\label{hybwfb}
\end{equation}
which may have different bipolaronic forms S=(S0), (S1$_x$), (S1$_y$),
(QS) each of which can be
small and large, so that in principle there are
8 different states. However, we have
not use
simultaneously all these states since there are no situations in the
parameter space $t,U$
where all their energies are simultaneously degenerate but only a
relevant subset
\footnote{Actually using all of them practically does not
change the result because the
irrelevant states hardly hybridize with the others.}.

Then the energy of the ground-state $E(K)$ has the
following variational form:
\begin{eqnarray}
<\Omega^{B}(K)|H|\Omega^{B}(K)>=
\frac{\sum_{S,S'} \beta_{S}\beta_{S'}^{*} M_{S,S'}}
{\sum_{S,S'} \beta_{S}\beta_{S'}^{*} P_{S,S'}},
	\label{HTEAbip}
\end{eqnarray}
where
\begin{eqnarray}
M_{S,S'}(K)&=&\sum_{p} e^{i K p}<\Psi_{S}^{B}(j)|H |
\Psi_{S'}^{B}(j+p)>\nonumber \\
	\label{ebip}
\end{eqnarray}
and
\begin{eqnarray}
P_{S,S'}(K)&=&\sum_{p} e^{i K p}<\Psi_{S}^{B}(j)|
\Psi_{S'}^{B}(j+p)>.\nonumber \\
	\label{scbip}
\end{eqnarray}

The extremalization of (\ref{HTEAbip}) with respect to  $\beta_{S}$ is done
by a diagonalization of the matrix $P^{-1/2}(K) M(K) P^{-1/2}$ of size $n
\times n$. The variational energy is minimized
with respect to  parameters of eq.(\ref{hybwfb}).

In all regions of the phase diagram,
for small $\alpha$
the HTEA energy corrections
for the bipolarons
(S0),(S1) or (QS) are systematically much smaller than those
 involved by the polarons.
The tunnelling energy
of bipolarons is much smaller than those of the polaron.

The hybridization cross-overs which are found at each smoothed
first-order transition of the TEA
phase diagram remain very narrow and the hybridization energy gain is
negligible.
One needs to have a high bipolaronic degeneracy such as the
triple point
or a  relatively large value of $\alpha$ ($\alpha \geq 0.05$)
to observe non negligible crossovers. Even in that case the
energy gains
remain small compared to those of an unbound pair of the HTEA polarons.

If the HTEA bipolarons keep almost the same energy as the TEA
bipolarons, the phase diagram fig.\ref{fig9} is practically unchanged. Of
course,
the first-order transition lines which appear in this phase diagram should
now be viewed as sharp crossover lines.
The crossover between the bound bipolarons and the unbound pair of polarons
has been investigated with a general HTEA bipolaron form
(including the latest)
but no significant hybridization has been found
between these two kinds of states so that we can not draw a conclusion
about the nature of this transition.

The triple point is a special point of the phase diagram where the bipolarons
(S0), (S1) and (QS) are degenerate
and where we should expect a higher energy gain by hybridization when
$\alpha$ is not too small.
Unfortunately, this triple point disappears when $\alpha$
increases beyond
approximately $0.002$. When it just disappears
the TEA bipolaron binding energy referred
to two unbound HTEA polaron is just zero but then its tunnelling energy
$T_b$ is
maximum (but still only $10^{-7} \times t$ : that is the bipolaron effective
mass is seven order
of magnitude larger than those of the bare electron).

The negative conclusion of this section is that more sophisticated variational
calculations does not confirm the conclusion of section (\ref{BBW}) which
was based on the assumption $\alpha$ small
extrapolated to larger $\alpha$.

The present study also shows that
in the domain of  small $U$ one may have a quantum
bipolaron ground-state with a large tunneling energy
occurring at very large $\alpha>0.1$.
This result is simply obtained with only the TEA of the
small bipolaron (S0) that is proved to have a nonnegligible
binding energy for both $t$ and $U$ small enough.
Nevertheless, this result is not relevant for such large $\alpha$,
our approach based on a perturbative theory of the adiabatic limit
fails because of too large quantum lattice fluctuations.

%%%%%%%%%%%%%%%%%%%%%%%%%%%%%%%%%%%%%%%%%%%%%%%%%%%%
\section{Phonon Dispersion Effect}
%%%%%%%%%%%%%%%%%%%%%%%%%%%%%%%%%%%%%%%%%%%%%%%%%%%%
\label{sect6}

We intend to show that highly degenerate point
that could persist
under large quantum lattice fluctuations implies very light bipolarons.
To achieve that goald, a simple procedure consists in changing
the model so as to favor the bipolaron (QS).
If we could make it more robust to quantum
lattice fluctuations it should become very light
for reasonably large $\alpha$ by
hybridization with the other degenerate bipolarons at the triple point.

We choose to introduce a phonon
dispersion, but this might not be the unique way.
When an electron is present at a given site it will
also distort the lattice at the neighboring sites. If the sign of
the dispersion is appropriate, the lattice potential at the neighboring sites
is lower which favors its occupancy by electrons and thus the spatial
extension of the bipolaron.
The bipolaron (QS) which is more extended than the bipolaron (S0)
should be favored.
\footnote{Phonon dispersion may induce other important effects in
the bipolaron structure as shown in \cite{RA95} for CDW's.}

We consider the new Hamiltonian
\begin{equation}
\mathcal{H}_d= \mathcal{H} -c\sum_{<i,j>} (a^+_i +a_i)(a^+_j +a_j)
\label{Hamdisp}
\end{equation}
where $\mathcal{H}$ is the Holstein-Hubbard Hamiltonian (\ref{hamiltonian})
and its reduced Hamiltonian corresponds to $H$ (\ref{hamduc})
that gives
\begin{equation}
H_d=H-\frac{C}{4} \sum_{<i,j>} u_i u_j
\label{handsduc}
\end{equation}
with
\begin{eqnarray}
C&=& \frac{4c}{E_0} (\frac{4g}{\hbar \omega_0})^2 \label{CC}\\
\end{eqnarray}

When the coupling $C$ is positive the dispersive term generates an effective
attractive interaction between polarons. This coupling cannot
exceed the value $1/2$ beyond which the low wavevector
phonons becomes unstable.

\subsection{Adiabatic Limit}

At the adiabatic limit the equation eq.\ref{densMF} becomes
\begin{equation}
\langle u_{i}\rangle =-\frac{1}{2} \sum_j D_{i,j}^{-1} \langle n_{j}\rangle
	\label{densMFd}
\end{equation}
where $D$ is the matrix :
\begin{eqnarray}
D_{i,i}&=&1\nonumber\\
D_{i,i \pm 1_x}=D_{i,i \pm 1_y}&=& -\frac{c}{2}   \nonumber\\
D_{i,j}&=&0 \qquad \mbox{otherwise}  \nonumber\\
\label{dispMatrix}
\end{eqnarray}

\begin{figure}
\begin{center}
\includegraphics{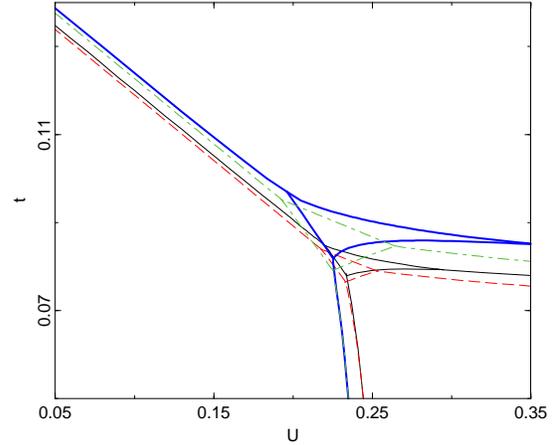}
\caption{\label{diagC1}
ground-state phase diagram for Hamiltonian (\ref{handsduc})
at $C=0.1$ (thick full lines) compared to the initial case $C=0$
( thin full lines),
and approximate diagrams calculated with the bipolaron exponential ansatz
(thin dashed lines)  for same couplings  $C=0.$, $C=0.1$.}
\end{center}
\end{figure}

Bipolarons (S0),(S1), (QS)... which were found at the anti-integrable
limit of the Holstein-Hubbard model at $t=0$ persist as ground-states in
this model with nonzero coupling $C$ \cite{PA99}
(see the diagrams figs.\ref{diagC1} and \ref{diagC3}).
The domain where bipolaron (QS) is the ground-state enlarges
when $C$ increases up to its maximum value $1/2$.
As expected the existence of a positive dispersion favors
the quadrisinglet ground-state.

The first-order transition between bipolaron (S0) and
(QS) becomes almost second-order and difficult to distinguish numerically
since there is no symmetry breaking between these two bipolaronic states.
Then as expected, there is a soft internal mode which almost vanishes at
the transition on both side of the transition
which corresponds to a breathing mode of the bipolaron with the same
symmetry. Simultaneously the Peierls-Nabarro barrier almost vanishes.

\begin{figure}
\begin{center}
\includegraphics[width=0.45 \textwidth]{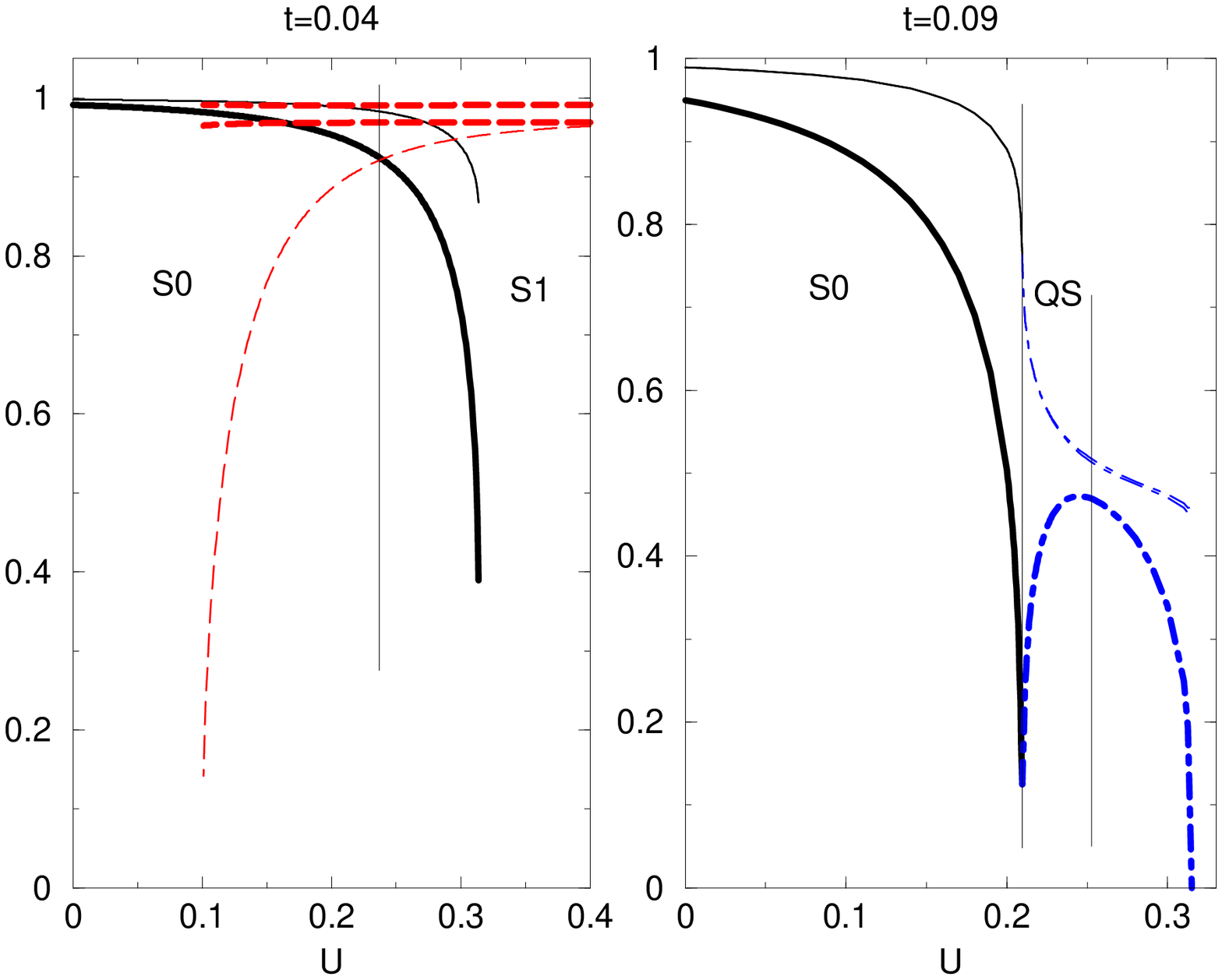}
\caption{\label{phonC} Frequencies versus $U$  of the internal modes
of bipolaron (S0) (full lines), (S1) (dashed lines), (QS) (dot-dashed lines),
for $t=0.04$ (left) and $t=0.09$ (right).
The breathing modes are represented by thick lines
and the pinning modes  by thin lines.
Vertical lines determine the location of the first order transitions.}
\end{center}
\end{figure}

\begin{figure}
\begin{center}
\includegraphics{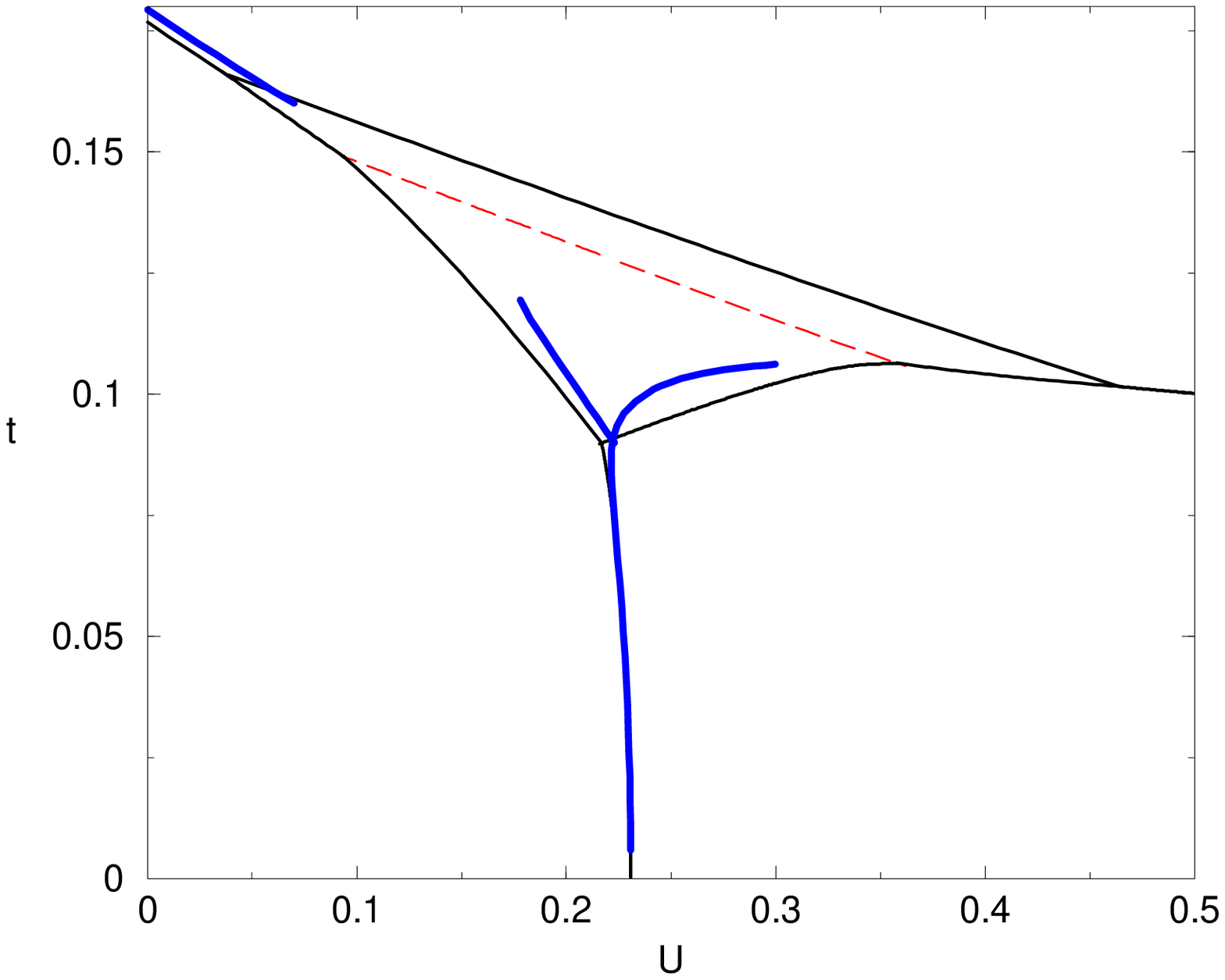}
\caption{\label{diagC3}
Same as fig.\ref{fig9} but with a phonon dispersion $C=0.3$
calculated exactly in the adiabatic limit (thick full lines)
and approximated  with the exponential ansatz (thin full lines).}
\end{center}
\end{figure}

This soft mode which does not break the bipolaron symmetry is not a pinning
mode and does not favor the classical mobility of this bipolaron.
To that purpose the pinning mode which also softens at the first order
transition
between (QS) and  (S1) is the most appropriate ( see fig.\ref{phonC} and
refs. \cite{PA98},\cite{PA99}).

\begin{figure*}
\begin{center}
\includegraphics[width=0.3 \textwidth]{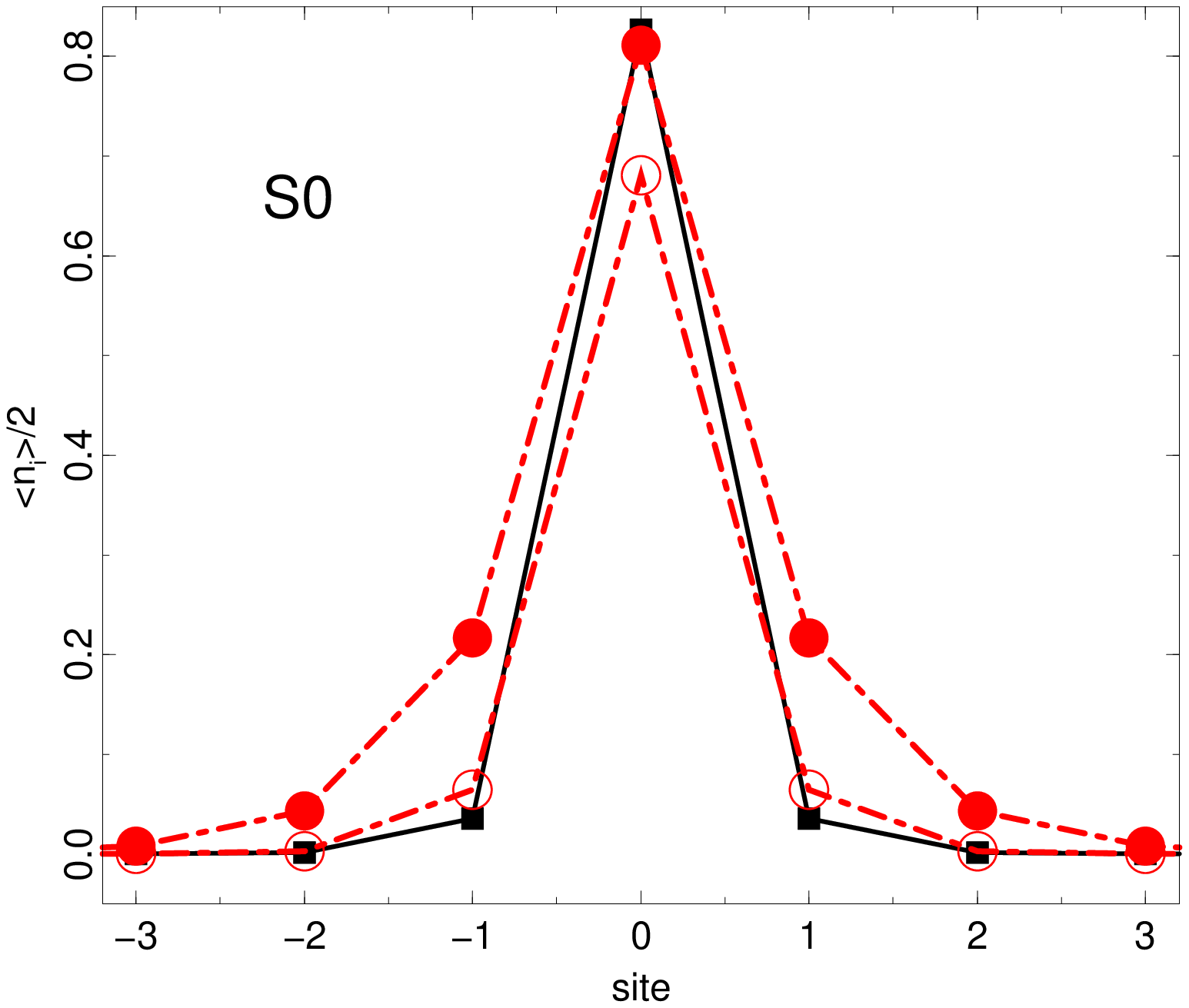}
\includegraphics[width=0.3 \textwidth]{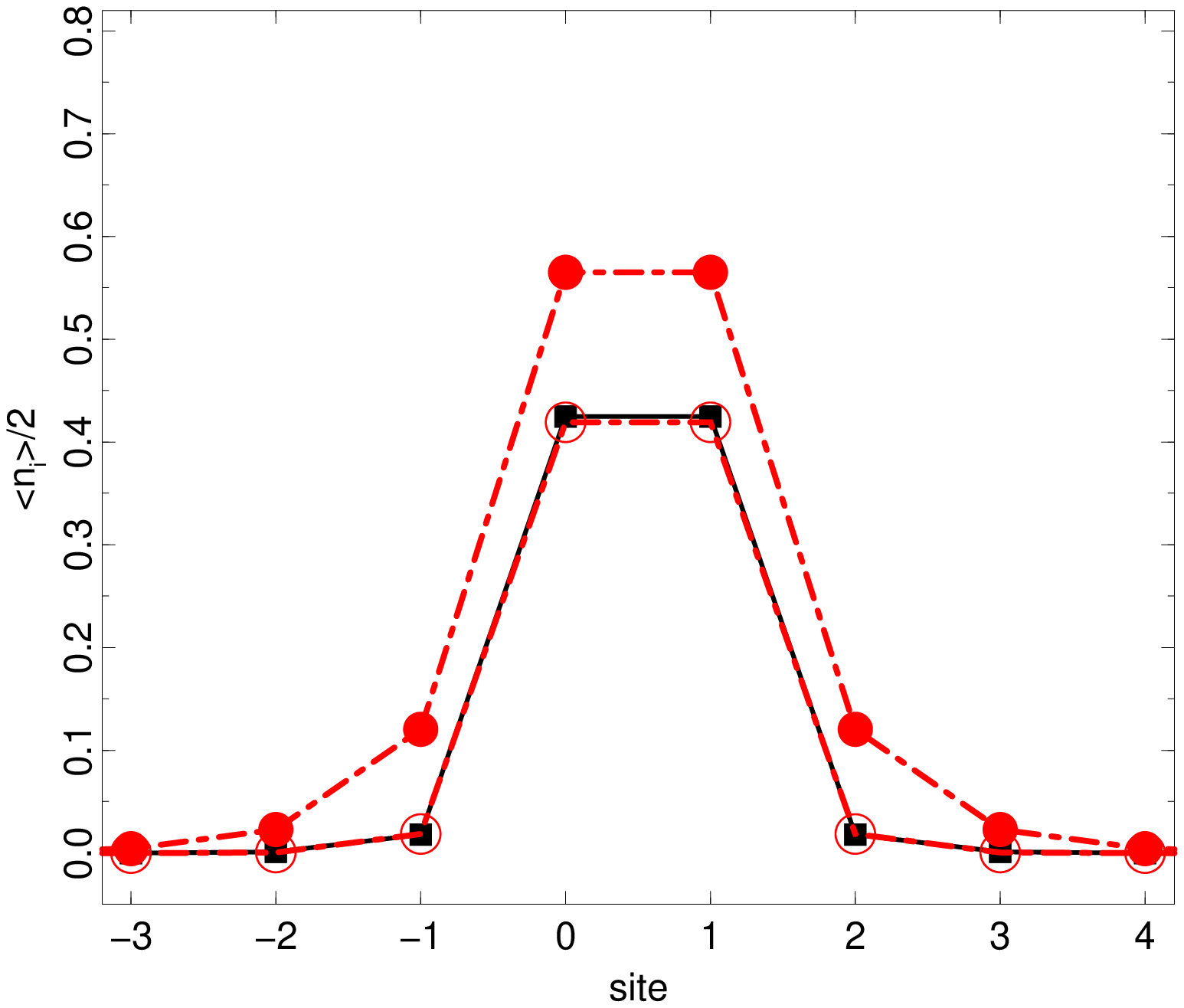}
\includegraphics[width=0.3 \textwidth]{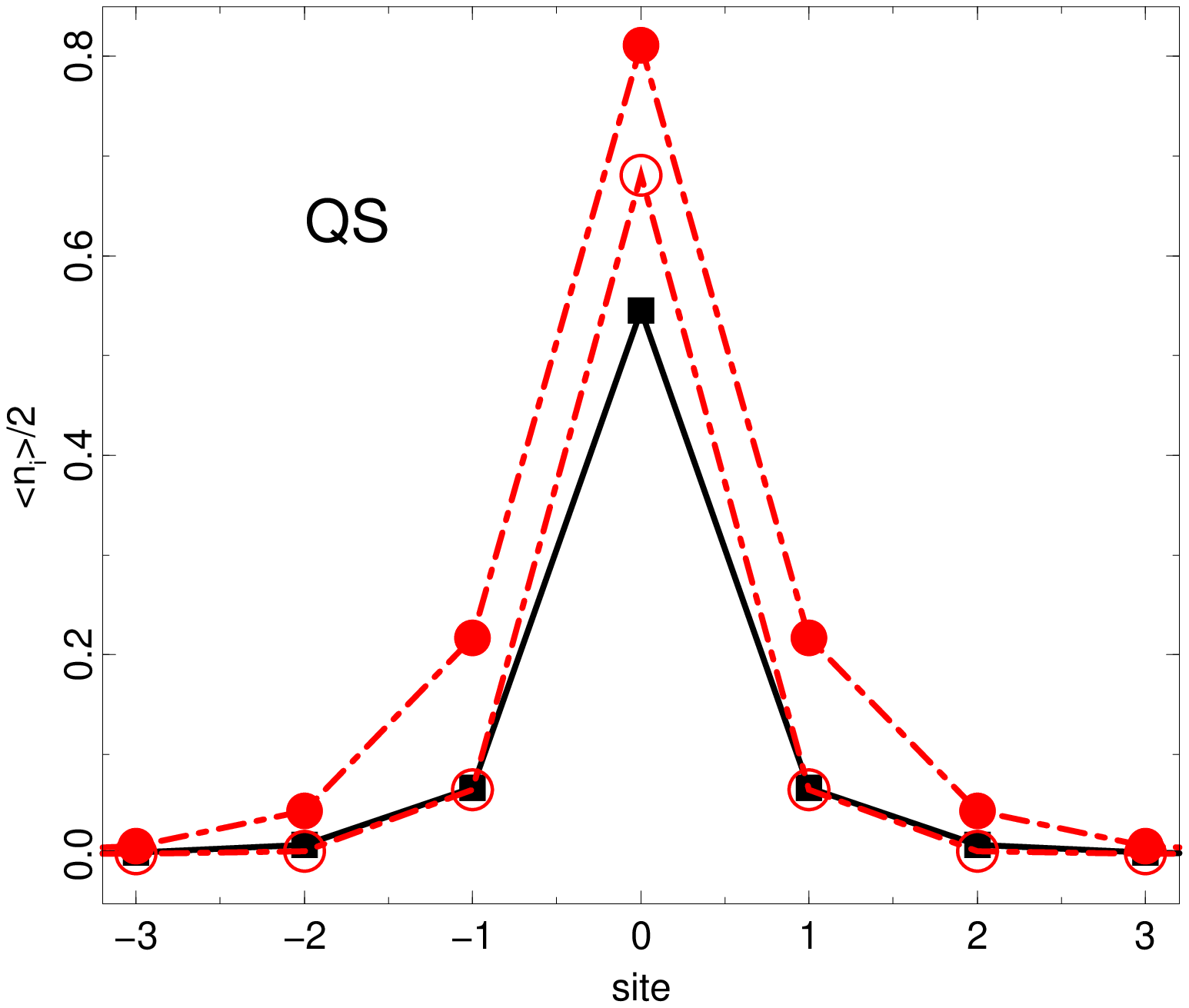}
\caption{\label{fig13}
Profiles of electronic density (empty symbol)
and absolute value of the displacement (full symbol)
versus site in x direction for the adiabatic bipolarons (S0),(S1) and (QS)
at the triple points: $C=0.$ $t=0.078$ $U=0.233$ (squares linked by 
full lines),
and $C=0.3$ $t=0.0904$ $U=0.222$ (circles linked by dot-dashed lines).}
\end{center}
\end{figure*}

When the coupling $C$ is too large $0.2<C<0.5$
our accuracy is  limited in practice because of the bipolaron (QS) extension,
which requires large system sizes we cannot afford.
This problem occurs especially close to the first-order transition
between bipolaron (QS) and the  extended state (see diagram
fig.\ref{diagC3}).
However the exponential ansatz eq.\ref{AS0},\ref{AS1},\ref{ASQ}
still fits quite well the bipolaron ground-state
as shown on the diag.\ref{diagC1}.
These variational forms allow us to compute quickly the bipolarons
even for large $C$ values and to determine approximately the ground-state
with a reasonable accuracy (see diagrams figs.\ref{diagC1} and \ref{diagC3}).

As we already know the flaw of this approximate method  is that
spurious first-order transitions may occur. This situation happens nearby the
first-order transition between (QS) and the extended state
as seen on diag.\ref{diagC3}. It is due to the exponential ansatz
which does not provide a good fit of the bipolaron  when it becomes
more extended.

However, at the triple point the bipolaron ground-state is
still localized on very few sites (fig.\ref{fig13}) for $C=0.3$ and
the exponential ansatz remains sufficiently accurate.

\subsection{Quantum Corrections}

\begin{figure*}
\begin{center}
\includegraphics[width=0.4\textwidth]{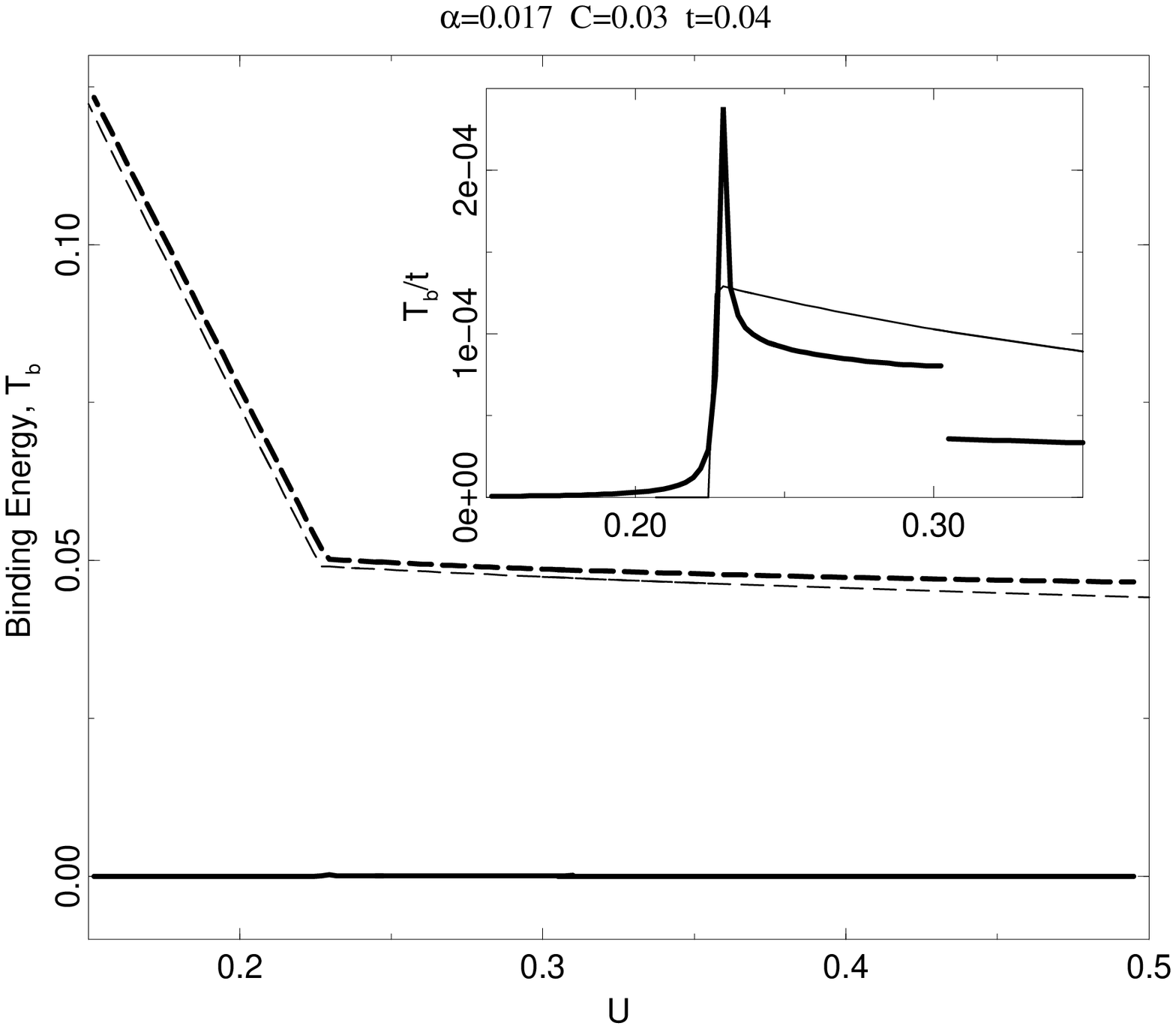}
\includegraphics[width=0.4\textwidth,height=0.35\textwidth]{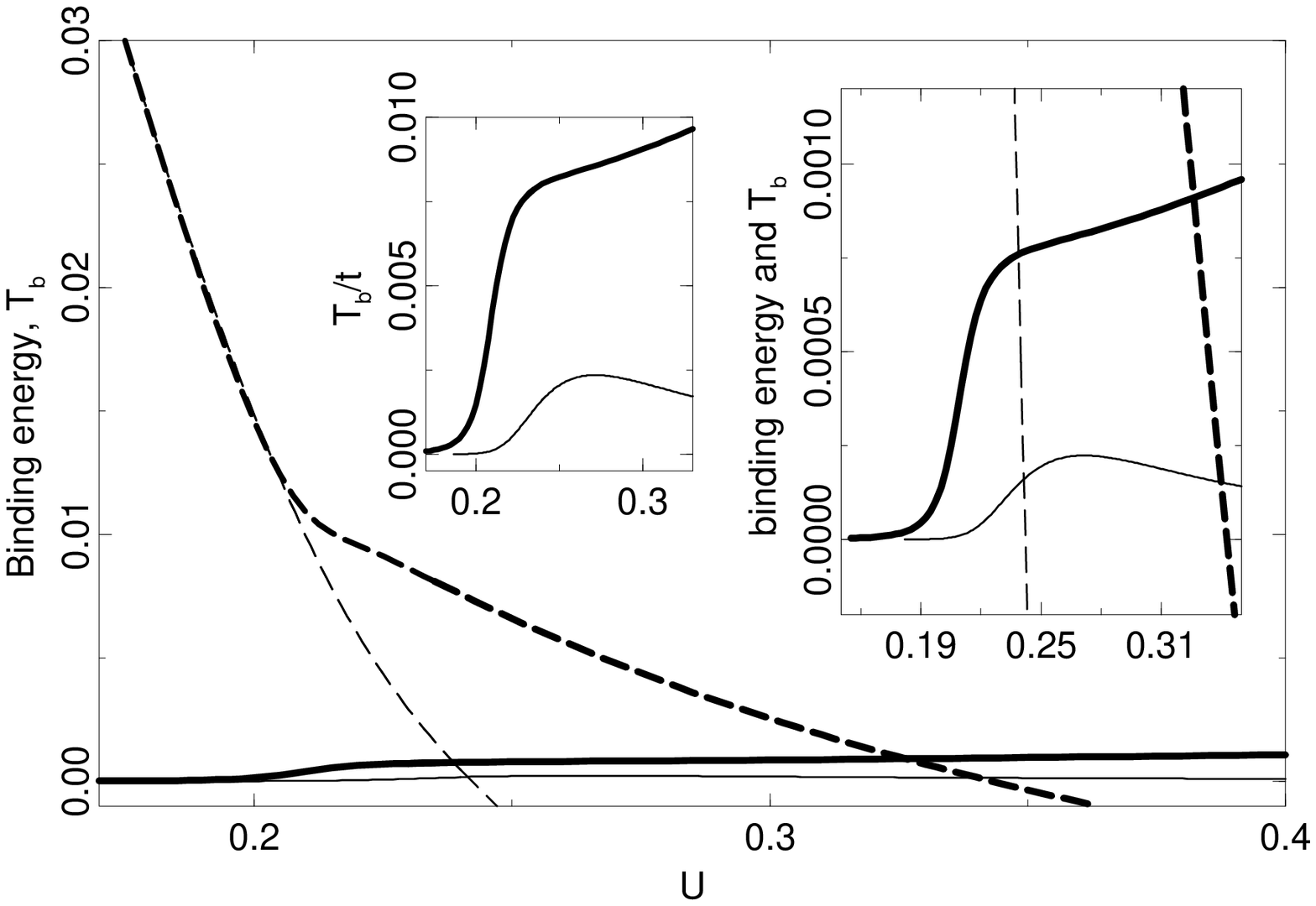}
\caption{Binding energy (dashed lines) and tunnelling energy $T_b$ versus
$U$ for the quantum bipolaron
ground-state at $t=0.04$ (left) and $t=0.095$ (right) for $C=0.3$ and
$\alpha=0.017$
calculated by the perturbative method (thick lines) and the HTEA method
(thin lines).}
\label{fig14}
\end{center}\end{figure*}

Same methods, as those used above
for the original  Holstein-Hubbard model
are applied to deal with the quantum lattice
fluctuations of the modified model.
The degeneracy due to the translation invariance of the model
is lifted according to
standard perturbation theory. One gets a tight  binding model as
in section \ref{BBW} which yields  both binding
and tunnelling energies of the quantum ground-state. Figs.\ref{fig14})
shows these quantities for a strong coupling  ($C=0.3$).

The binding energy of the bipolaron refers to two non-interacting polarons
calculated with the HTEA method, which is the most accurate.
For a single polaron condition (\ref{denposf}) becomes

\begin{equation}
    \sum_{i} v_{i}^P= - \frac{1}{2(1.-2.*C)}
	\label{denposfD}
\end{equation}
and we choose to write the displacement as $v^P=D^{-1} v$
where $v$ is given by
\begin{eqnarray}
v_{i}&=& -B \mu^{|i_x|+|i_y|} \label{VpolD}
\end{eqnarray}

For a large enough phonon coupling $C>0.2$, in the region
we investigate $t<0.1$ the HTEA method only requires the
hybridization  between a small polaron and a large polaron.
The almost second-order transition displayed by
the TEA at $C=0$ occurs now at a larger $t_p^2(\alpha)$.

The binding energy of the
quantum bipolaron is still large in that region and one notices the
optimal regime where
both tunnelling and binding energies have the same value.

To obtain the optimal region,  a fine tuning of the parameters
is required
because changing them slightly can  either reduce the
binding energy so that the bipolaron becomes fragile against temperature  or
sharply increase its effective mass, killing its quantum mobility.

Phonon dispersion favors the mobility of the bipolaron because it
extends the lattice distortion  around the bipolaron
(see fig.\ref{fig13}) as well as the electronic wave function.
Classically, this effect is manifested by internal mode softening
and by the depression of the Peierls-Nabarro energy barrier (not calculated
here see paper I \cite{PA99}) between the different bipolarons. As a result,
when the lattice is quantum the hybridization between
the different bipolarons is increased, which increases the band width
and decreases the effective mass.

The HTEA calculation  (\ref{hybwfb}) for the
bipolaron confirms these  properties
(see figs.\ref{fig14}). Condition (\ref{denbisf}) becomes

\begin{equation}
    \sum_{n} v_{n}^{B}= - \frac{1}{(1.-2.*C)}
	\label{denbisfD}
\end{equation}
and $v^B=D^{-1} v$ where $v$ is still given by eq.(\ref{VpolD}).

In the vicinity of the (QS) region (see
figs.\ref{fig14})
the effective mass of the HTEA for the bipolaron is about five times larger
than the effective mass computed with the perturbative method, but the
bipolaron mass is still very small.
The comparison of the binding energy calculated with
the two methods shows that the variational HTEA method is not accurate in
the area
of the QS region. Indeed the perturbative method gives a stronger binding
energy and
thus it is variationally better. This is likely due to the fact that
when the bipolaron extends too much the TEA is not accurate because
the bipolaron shape is not well approximated by the exponential.

Fig.\ref{fig14} shows for $\alpha=0.017$ the effective
mass of the bipolaron in the optimal regime that ranges not far from 100
bare electronic mass. We choose as an example the realistic optical phonon
frequency $\hbar \omega_0=1.10^{-1} eV$ and to be in the optimal regime
$C=0.3$, $\alpha=0.017$ $U=0.25$ $t=0.1$ the initial parameter of
Hamiltonian (\ref{hamiltonian}) must be $g=3.10^{-2} eV$ $E_0=6 eV$
$\upsilon=1.5 eV$ $t=0.6 eV$ $c=0.3 eV$.
The tunnelling energy as well as the bipolaron binding
energy are about $6.10^{-3} eV$. With such characteristic values and
a  bipolaron  concentration not too large, a superfluid state could be
expected at relatively high temperatures,
perhaps few hundred degrees K. This estimate neglects the bipolaron
interactions, but when their concentration becomes large
these interactions
cannot be neglected, especially at half filling when there is one polaron
per site.
Close to this close packing regime the bipolaronic structure
cannot exist anymore for sure. Instead, a magnetic
spatially ordered polaronic structures could occur. Further studies
should investigate the
situation with large electron densities.

\section{Concluding Remarks}

In some circumstances the bipolaron
might become unusually light, which allows in principle the formation
of superconducting states at rather high temperature with physically
realistic parameters.
This effect is due to the degeneracy of several
bipolaronic states in the adiabatic limit
for some specific regions of the phase diagram.
In this situation there are small Peierls Nabarro barriers
and phonon softening for the different bipolaronic states.
Then the quantum lattice fluctuations lift the degeneracy
between the degenerate states and
may yield very light hybridized bipolarons, which however
are well-bound.

We realized this situation in a modified Hols\-tein-Hub\-bard model,
which involves both an electron-phonon interaction and a direct repulsive
electron-electron interaction.

The superconducting state of such very light bipolarons occurs for
weak concentrations. When the concentration becomes larger there are
strong interactions between the bipolarons,
 which may both break them into polarons
and organize different structures (for example, magnetic).

This situation may happen in superconducting cuprates.
In the undoped regime where the band of electrons is half filled,
the structure can be viewed as close-packed polarons with
an antiferromagnetic ordering. This polaron structure should persist
for low doping till a certain electron concentration where the holes
are polaron vacancies.
For a sufficiently large doping the electron concentration may
become low enough in order that a (first order)
transition toward a superfluid of light quantum  bipolarons
takes place.
The real phenomenology should  be more complex because one should expect
that the model parameters depend on the doping and thus that the system
does not remain always close from the optimal regime with strongly bonded
light bipolarons but move around this point.
Otherwise, we suggested in \cite{PA99} that in some appropriate  models
the (QS) bipolaron could have a d-symmetry. We have not yet realized
an explicit model where such an effect occurs, but we hope to.

The numerical techniques we used (Toyozawa Exponential Ansatz) and
its improvement (HTEA) where the hybridization between different
states is taken into account,
turned out to be very efficient to study the bipolaron mass.
It should be developed  to consider models with many electrons.
In \cite{AAR92} it was proven that at adiabatic limit, the ground-state
at large electron-phonon coupling was bipolaronic.
Variation of the exponential ansatz may provide strong simplifications
for these case and a qualitative understanding of the many-polaron problem
first in the adiabatic limit, next with quantum lattice fluctuations.
Finally, the problem of quantization of discrete breathers can be approached
with similar techniques \cite{PA99b}.

%------------------------------------------------------

\end{document}